\documentclass[aps,prb,preprintnumbers,amsmath,amssymb,superscriptaddress,twocolumn,10pt]{revtex4-2}

\usepackage{txfonts}
\usepackage{graphicx}
\usepackage{dcolumn}
\usepackage{bm}
\usepackage{array}
\usepackage[dvipsnames]{xcolor}
\usepackage[%
    colorlinks=true,
    pdfborder={0 0 0},
    linkcolor=blue,
    citecolor=black
]{hyperref}  
\usepackage{chngpage}
\usepackage{appendix}

\usepackage{blindtext,subcaption,graphicx}
\usepackage{booktabs}
\usepackage{caption}
\usepackage[T1]{fontenc}
\usepackage{lmodern}

\definecolor{Dark}{gray}{0.2}
\definecolor{MedDark}{gray}{0.4}
\definecolor{Medium}{gray}{0.6}
\definecolor{Light}{gray}{0.8}
\definecolor{darkred}{rgb}{0.55, 0.0, 0.0}
\definecolor{darkslateblue}{rgb}{0.28, 0.24, 0.55}
\definecolor{royalblue(web)}{rgb}{0.25, 0.41, 0.88}
\usepackage{graphicx,float}
\usepackage{amsmath}

\usepackage{amssymb}
\hypersetup{
    citecolor=darkred,
    linkcolor=blue,   
    urlcolor=blue}

\usepackage{bbm}

\def\A{\mathrm{A}}
\def\B{\mathrm{B}}

\def\be{\begin{equation}}
\def\ee{\end{equation}}
\def\beq{\begin{equation}}
\def\eeq{\end{equation}}
\def\bea{\begin{eqnarray}}
\def\eea{\end{eqnarray}}
\def\nbea{\begin{eqnarray*}}
\def\neea{\nonumber\end{eqnarray*}}
\def\bmat#1{\left(\begin{array}{#1}}
\def\emat{\end{array}\right)}
\def\bcase#1{\left\{\begin{array}{#1}}
\def\ecase{\end{array}\right.}
\def\bmini#1{\begin{minipage}{#1\textwidth}}
\def\emini{\end{minipage}}

\usepackage{wrapfig}
\graphicspath{{figures/}}
\usepackage{enumerate}

\begin{document}

\title{Field-Induced Ordered Phases in Anisotropic Spin-1/2 Kitaev Chains}
\date{\today}
\author{Mandev Bhullar}
\affiliation{Department of Physics, University of Toronto, 60 St. George St., Toronto, Ontario, Canada M5S 1A7}
\author{Haoting Xu}
\affiliation{Department of Physics, University of Toronto, 60 St. George St., Toronto, Ontario, Canada M5S 1A7}
\author{Hae-Young Kee}
\email[]{hy.kee@utoronto.ca}
\affiliation{Department of Physics, University of Toronto, 60 St. George St., Toronto, Ontario, Canada M5S 1A7}
\affiliation{Canadian Institute for Advanced Research, CIFAR Program in Quantum Materials, Toronto, Ontario, Canada M5G 1M1 }

\begin{abstract}
Motivated by intense research on two-dimensional spin-1/2 Kitaev materials, Kitaev spin chains and ladders, though geometrically limited, have been studied for their numerical simplicity and insights into extended Kitaev models. The phase diagrams under the magnetic field were also explored for these quasi-one dimensional models. For an isotropic Kitaev chain, it was found that a magnetic field polarizes the ground state except along the symmetric field angle, where the chain is found to remain gapless up to a critical field strength where it enters an intriguing soliton phase before reaching the polarized state at higher field strengths. Here we study an anisotropic Kitaev chain under a magnetic field using the density matrix renormalization group technique, where the ground state has a macroscopic degeneracy with a finite gap in the absence of the magnetic field. 
When the field is mainly aligned parallel to the strong bond, four-site and large unit-cell ordered phases arise. In a certain angle of the field, another ordered phase characterized by a uniform chirality with six-site periodicity emerges. We employ a perturbation theory to understand such field-induced ordered phases. The effective model uncovers the presence of transverse Ising and Dzyaloshinskii-Moriya interactions between unit cells, as well as further-neighbor Ising interaction induced by the magnetic field, which collectively explain the mechanisms behind these ordered states.
Open questions and challenges are also discussed.
\end{abstract}

\maketitle

\section{Introduction\label{intro}}
The spin-1/2 Kitaev model with bond-dependent Ising interaction on a honeycomb lattice has attracted significant interest, as it offers a rare example of a quantum spin liquid ground state \cite{Kitaev2006}.
Furthermore, the Jackeli-Khaliullin mechanism to generate the bond-dependent interaction utilizing spin-orbit coupling \cite{Jackeli2009PRL} has inspired numerous extended studies \cite{Rau2016ARCMP,Winter2016PRB,Winter2017,Hermann2018PRB,Takagi2019NRP,Motome2019JPSJ,Takayama2021JPSJ,Trebst2022,Rouso2024RoPP}. A few examples of early studies include the generic spin model with the off-diagonal symmetric $\Gamma$ interaction \cite{Rau2014} and candidate material proposals such as A$_2$IrO$_3$ \cite{Singh2010PRB,Singh2012PRL,Liu2011PRB} and $\alpha$-RuCl$_3$ \cite{Plumb2014PRB,Kim2015}. 
The one-dimensional (1D) counterparts such as chains and ladders of two-dimensional (2D) extended Kitaev models have been also investigated due to their numerical accessibility \cite{CatuneanuPRB2019,Yang2020PRL,Sorensen2021PRX,Rouso2024RoPP}.
Despite their limited geometry, they have offered some insight into the 2D models. For example, magnetically disordered states found in the 2D models using numerical studies are found to also be disordered in the chain and ladder geometries, signaling intriguing nature of the disordered states in the 2D limit \cite{Gordon2019NC, Sorensen2021PRX}. 

The effect of the magnetic field on the Kitaev chain and ladder has been another direction that attracted many studies \cite{Sun2009,Wang2010,You2018,Wu2019,Sorensen2021PRX}. Most studies were further motivated by unconventional behavior of the thermal Hall conductivity observed near the transition from the magnetic ordered state to a field-induced phase reported in $\alpha$-RuCl$_3$ \cite{kasahara2018thermal}. 
The field-induced intermediate phases in the extended Kitaev ladder and 24-site honeycomb lattice model under the [111] field direction \cite{Gordon2019NC} have spawned further numerical studies \cite{Lee2020NC,Gohlke2020PRR,Liu2021PRR}.

Focusing on the spin-1/2 Kitaev chain, it was shown in earlier studies that the model under a transverse magnetic field is exactly solvable and a direct transition to the polarized state at any finite field was found \cite{Sun2009}. However, under longitudinal fields, the exact solvability is broken, and a recent numerical study on the isotropic chain found the polarized state in most phase space of the magnetic field except that applied along the symmetric field angle,
where an interesting chiral soliton phase emerges \cite{Sorensen2023PRRa}.
Unlike the isotropic spin-1/2 Kitaev chain which has gapless excitations, the anisotropic spin-1/2 Kitaev chain has a microscopic degeneracy with a finite gap. Thus, the effects of the magnetic field on the anisotropic chain are expected to be widely different. 

Here, we investigate responses of the anisotropic spin-1/2 Kitaev chain  
to longitudinal magnetic fields, and report field-induced ordered phases of four sites and six sites with uniform chirality. We employ a perturbation theory to explain the mechanism behind such field-induced orders. The low-energy model includes an effective magnetic field in addition to transverse Ising and Dzyaloshinskii-Moriya interactions, which are the sources of the field-induced orders. Near the transition to the polarized state (PS), additional large unit-cell (\(\mathrm{LU}\)) phases are found, and higher-order exchange interactions are derived to understand the mechanism of the \(\mathrm{LU}\) phases. 

The paper is organized as follows. In Sec. \ref{review}, we briefly review the model and the solution of the ground state with the transverse field. In Sec. \ref{numerics}, we present the phase diagram under longitudinal fields obtained using the DMRG method. In Sec. \ref{perturbationtheorysection}, a perturbation theory up to third order is presented to explain the field-induced orders. In Sec. \ref{exactlysolvablepointsection}, exactly solvable points of the model are presented to explain the soliton phase. In Sec. \ref{luphasediscussion}, we present large unit-cell phases which require perturbation theory beyond third-order terms to explain.
A summary and discussion is given in the final Sec. \ref{summary}. 

\section{Review of spin-1/2 Kitaev chain under transverse magnetic field\label{review}}
In this section we review the model Hamiltonian and its exact solution under an external transverse magnetic field \cite{Sun2009}. The Hamiltonian of the anisotropic Kitaev chain under a magnetic field is 
\begin{equation}
\begin{aligned}
    \mathcal{H} = \sum_{j=1}^{N_c} \bigg[ \frac{K_x}{4} \sigma_{j,\A}^x \sigma_{j,\B}^x + \frac{K_y}{4} \sigma_{j,\B}^y \sigma_{j+1,\A}^y \\  +  \frac{1}{2} \mathbf{h} \cdot (\boldsymbol{\sigma}_{j,\A} + \boldsymbol{\sigma}_{j,\B})\bigg],  
    \label{FullHamiltonian}
\end{aligned}
\end{equation}
where \(j\) indexes one out of \(N_c\) unit cells, each containing two sublattice sites \(A\) and \(B\). When \(\mathbf{h} = (0, 0, h_z)\), the Hamiltonian can be written as a quadratic form in terms of Majorana fermions through a Jordan-Wigner transformation. There are four energy bands for the Majorana fermions, $(\epsilon_{k,+},\epsilon_{k,-},-\epsilon_{k,+},-\epsilon_{k,-})$, where 
\begin{equation}
\begin{aligned}
    \epsilon_{k,\pm} =& \frac{1}{2} \left( \sqrt{K_x^2 + K_y^2 +2 K_x K_y \cos k + 16 h_z ^2 }\right. \\
    &\pm  \left.  \sqrt{K_x^2 + K_y^2 + 2K_x K_y \cos k }\right). 
        \label{MajoranaFermionsEnergyBands}
\end{aligned}
\end{equation}
The ground state of the spin model corresponds to the fermion model with zero Fermi energy. When \(h_z=0\), \(\pm \epsilon_{k,-}\) become flat and vanish. This indicates that there is a \(2^{N_c}\)-fold degeneracy for every eigenstate of the system \cite{Wang2010, You2018,Wu2019, feng2007topological}.  

We are now interested in the effect of the longitudinal field, \(\mathbf{h} = (h_x, h_y, 0)\). A finite longitudinal field introduces interaction terms for the Majorana fermions, and with the exception of certain special points (as we will see), the exact solvability of the model breaks down. 
The effect of the longitudinal field on the isotropic chain \(K_x = K_y\) has previously been studied numerically in Ref. \cite{IsotropicKitaevUnderFieldPaper}, and several interesting phenomena were observed. In particular, for a field applied along the diagonal \(h_x = h_y\), the chain is observed to remain gapless up to a critical field strength where it enters an intermediate soliton phase \(\mathcal{B}\) before reaching the polarized state at higher fields, as shown in Fig. \ref{Alpha15PhaseDiagram}. The \(\mathcal{B}\) phase is characterized by a staggered vector chirality and for periodic boundary conditions (PBCs), a two-fold degenerate ground state with a finite gap. 

In this paper, through a combination of both numerical and analytical techniques, we show that the longitudinal field can induce several intermediate phases in the spin-\(1/2\) anisotropic Kitaev chain. One of these phases is the \(\mathcal{B}\) phase as observed in the isotropic chain. Of greatest interest, however, is the appearance of several ordered phases. We first provide a numerical study on these various observed phases, and we proceed to conduct analytical investigations regarding their underlying nature. 

\section{Phase diagram obtained by DMRG method\label{numerics}}
In what follows, we parameterize \((K_x, K_y) = (K \cos (\alpha), K \sin (\alpha))\) so that \(K\) and \(\alpha\) control the bond strength and anisotropy. We set \(K = \sqrt{2}\) so that at the isotropic point with \(\alpha = 45^{\circ}\), \(K_x = K_y = 1\), which was the choice of parameterization used in Ref. \cite{IsotropicKitaevUnderFieldPaper}, and we may then directly compare our results to the known results of the isotropic chain. Without loss of generality, we will assume \(K_x \geq K_y \geq 0\), which implies we take \(0 \leq \alpha \leq 45^{\circ}\). 

We will also set \((h_x, h_y) = (h \cos (\phi_{xy}), h \sin (\phi_{xy}))\) and we restrict to the upper right quadrant of the \((h_x, h_y)\) plane. Unless otherwise noted, we also assume periodic boundary conditions (PBCs). 

To study the phase diagram of the anisotropic chain,
we employ the finite density matrix renormalization group (DMRG) method \cite{white1992density, schollwock2011density} for a system size of \(N = 2N_c = 200\) sites using the ITensor library \cite{DMRGRef}. The method is typically performed with a bond dimension larger than \(1000\) and a cutoff \(\epsilon < 10^{-10}\). The boundaries separating different phases are obtained by determining the sharp maxima in the energy susceptibility \(\chi_{\phi_{xy}}^{e}=-\partial^2 e_0/\partial \phi_{xy}^2\) for fixed \(h\), where \(e_0\) is the energy per site. 

\begin{figure}[!ht] 
    \centering
        \centering    
        \begin{centering}        \includegraphics[width=1\columnwidth]{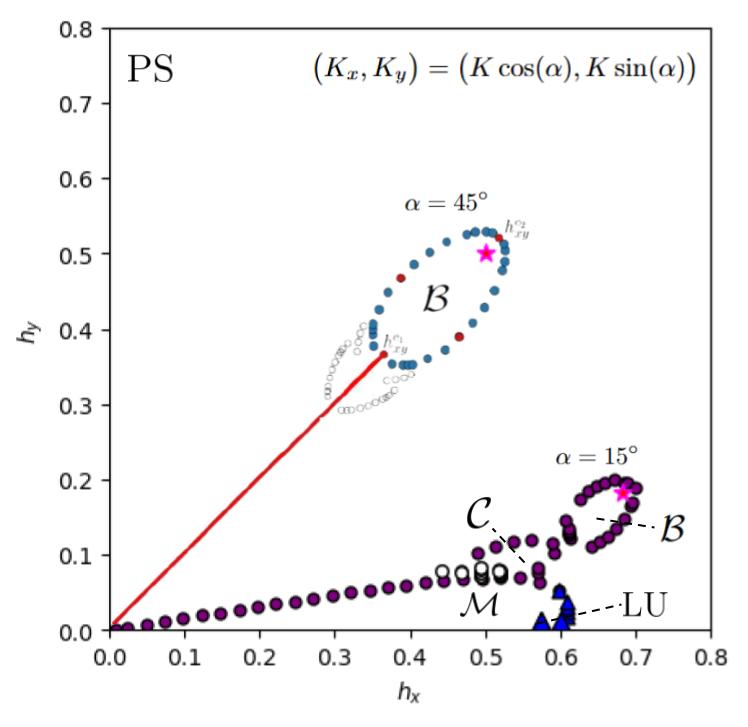}
        \end{centering}
  \caption{The phase diagram for the anisotropic chain for \(\alpha = 15^{\circ}\) in the \((h_x, h_y)\) plane, obtained from DMRG for \(N = 200\) sites, with PBCs (solid purple points). Blue triangles denote transition points from the \(\mathcal{M}\) phase to other smaller phases denoted by \(\mathrm{LU}\) occurring before the system reaches the \(\mathrm{PS}\). White circles denote a small region which separates the $\mathcal{M}$ and $\mathcal{C}$ phases (see the main text for details). For comparison, the phase diagram for the isotropic chain (\(\alpha = 45^{\circ}\)) obtained from Ref. \cite{IsotropicKitaevUnderFieldPaper} is also shown. Pink stars mark the exactly solvable points for the \(\mathcal{B}\) phase discussed in Sec. \ref{exactlysolvablepointsection}.}
\label{Alpha15PhaseDiagram}
\end{figure}

The result is shown for \(\alpha = 15^{\circ}\) in Fig. \ref{Alpha15PhaseDiagram}. The solid purple points forming the phase boundaries denote points where \(\chi_{\phi_{xy}}^{e}\) is sharply peaked. Our results indicate three major phases: \(\mathcal{M}\) and \(\mathcal{C}\) which are field-induced ordered phases as we will show below. The soliton phase \(\mathcal{B}\) identified in the isotropic chain is also observed.  
Blue triangles denote the transition from the \(\mathcal{M}\) phase to other \(\mathrm{LU}\) phases that occur before the system polarizes. 
White circles denote what is possibly incommensurability effects, which may indicate other tiny phases as well, and we leave the task of understanding this tiny area for future work, and focus on the nature of the major phases. The pink star denotes an exactly solvable point of the model contained within the \(\mathcal{B}\) phase, which will be discussed later. 

For comparison, we include in Fig. \ref{Alpha15PhaseDiagram} the phase diagram for the isotropic chain \(\alpha = 45^{\circ}\) which was previously obtained in Ref. \cite{IsotropicKitaevUnderFieldPaper} using Lanczos exact diagonalization (ED) \cite{weisse2008exact} for \(N = 24\) sites (solid blue points) in combintation with infinite DMRG (iDMRG) \cite{mcculloch2008infinite}  (solid red points). The gapless line is shown in red. The critical field strength for which the system transitions to the \(\mathcal{B}\) phase is found to be \(h_{xy}^{c_1} = 0.511\). The critical field strength for which the \(\mathcal{B}\) phase closes and transitions to the PS is found to be \(h_{xy}^{c_2} = 0.726 \). The area surrounded by the white circles denotes an area of incommensurability effects that was previously not well understood. An exactly solvable point is also found within the \(\mathcal{B}\) phase for the isotropic case.   

We show phase diagrams for other values of \(\alpha\), namely \(\alpha = 5^{\circ}\) and  \(30^{\circ}\) given in Fig. \ref{Alpha5Alpha30PhaseDiagram} in Appendix \ref{MoreNumericalDataSection}. As can be seen, the critical field at which the \(\mathcal{M}\) phase closes along the \(h_y = 0\) line decreases as \(\alpha \) increases, until the isotropic limit \(\alpha = 45^{\circ}\) is reached, where the \(\mathcal{M}\) is found to vanish over all phase space. It can also be seen that at lower field strengths, the critical field direction at which the \(\mathcal{M}\) phase transitions to the \(\mathrm{PS}\) increases as \(\alpha\) increases.

To study the characteristics of the chain in the \(\mathcal{M}\), \(\mathcal{C}\) and \(\mathcal{B}\) phases, we examine the on-site magnetizations \(\langle S_i^{x/y} \rangle \) for open boundary conditions (OBCs) in each of these phases in addition to the PS, which will aid us in understanding the behavior in PBCs as well. We show the results in Fig. \ref{Alpha15AllMagnetizations} below. Near the transition to the PS phase at small $h_y$, there is a small window of the LU phases, which will be discussed later in Sec. \ref{luphasediscussion}.

\begin{figure*}[!ht]
    \centering
    \captionsetup[subfigure]{justification=centering}
    \begin{subfigure}[b]{0.4\textwidth}         \centering         
    \caption{\(h = 0.55, \, \phi_{xy} = 4^{\circ}\)}
    \begin{centering}
    \includegraphics[width=0.95\textwidth]{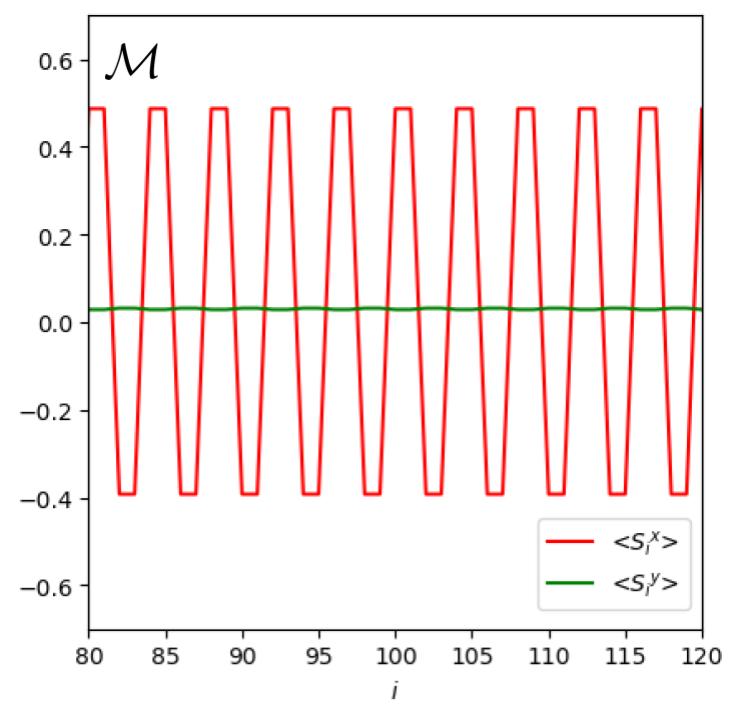} 
    \end{centering}
    \label{Alpha15MPhaseMagnetizations}
     \end{subfigure}
     \begin{subfigure}[b]{0.4\textwidth}
         \centering      
     \caption{\(h = 0.55, \, \phi_{xy} = 12^{\circ}\)}
     \begin{centering}
     \includegraphics[width=0.95\textwidth]{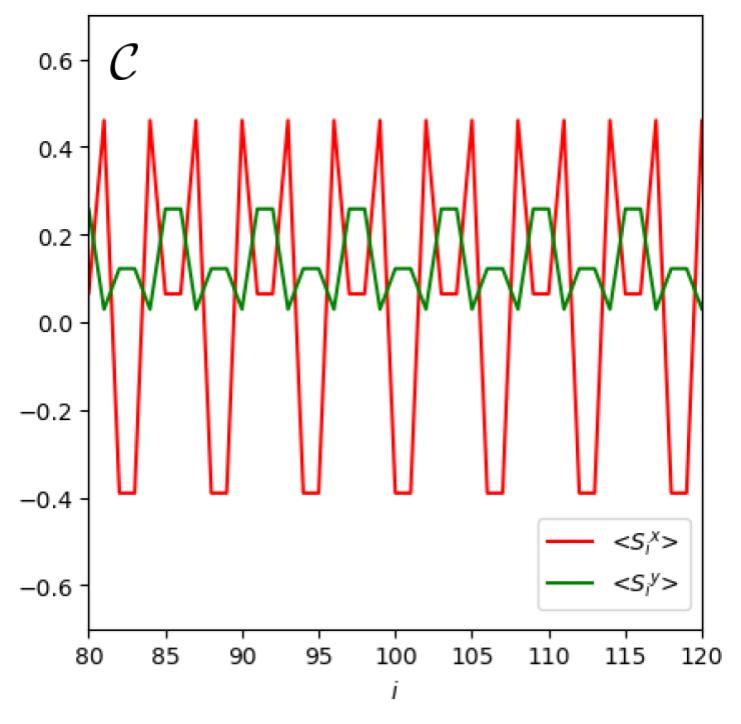}
     \end{centering}
       \label{Alpha15CPhaseMagnetizations}
     \end{subfigure}
     \begin{subfigure}[b]{0.4\textwidth}
         \centering        
    \caption{\(h = 0.65, \, \phi_{xy} = 12^{\circ}\)}
    \begin{centering}
    \includegraphics[width=0.97\textwidth]{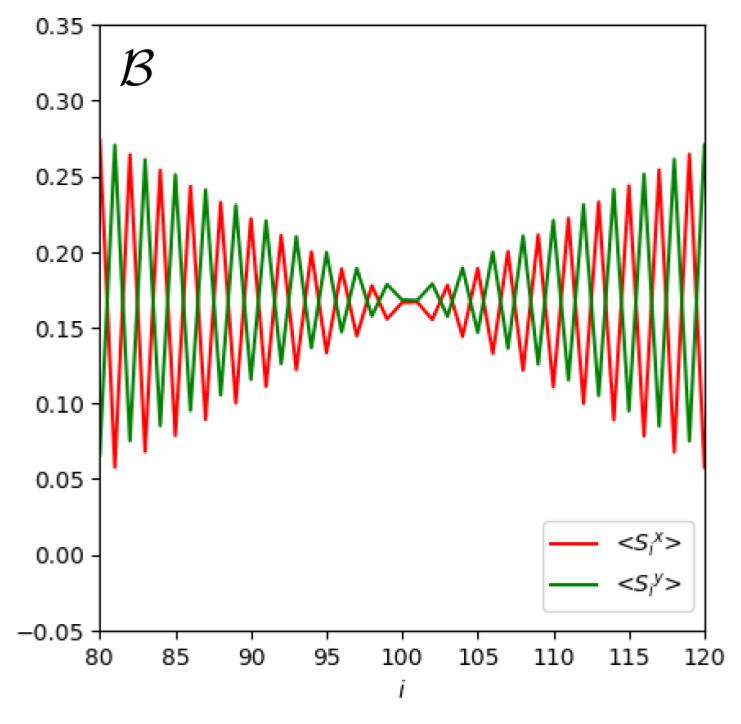}  
    \end{centering}     \label{Alpha15BPhaseMagnetizations}
     \end{subfigure}
     \begin{subfigure}[b]{0.4\textwidth}
         \centering      
    \caption{\(h = 0.65, \, \phi_{xy} = 20^{\circ}\)}
    \begin{centering}
    \includegraphics[width=0.95\textwidth]{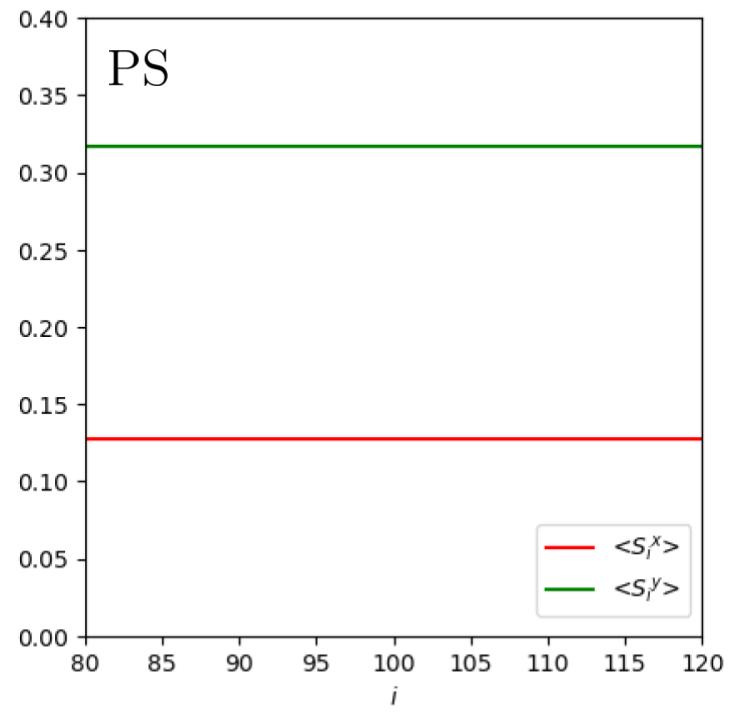} 
    \end{centering}
    \label{Alpha15PSMagnetizations}
     \end{subfigure}
  \caption{The on-site magnetizations \(\langle S_i^x \rangle\), \(\langle S_i^y \rangle\) versus position \(i\) in the middle of the chain for \(\alpha = 15^{\circ}\), obtained from \(N = 200\) site DMRG with OBCs, (a) within the \(\mathcal{M}\) phase, (b) within the \(\mathcal{C}\) phase, (c) within the \(\mathcal{B}\) phase and (d) in the PS.} 
\label{Alpha15AllMagnetizations}
\end{figure*}

From Fig. \ref{Alpha15MPhaseMagnetizations}, it is evident that the \(\mathcal{M}\) phase is an ordered phase that possesses a magnetic unit cell of 4 sites. In aligning the quantization axis along the \(x\)-axis, we would see that the ground state is two-fold degenerate, with the two spin configurations given by 
\begin{equation}
\begin{aligned} |\Psi_G^{\mathcal{M}} \rangle = \begin{Bmatrix}
    (|\uparrow \downarrow \downarrow \uparrow \rangle)^{\otimes N_c/2}, \\ (|\downarrow \uparrow \uparrow \downarrow \rangle)^{\otimes N_c/2}
\end{Bmatrix}.
\label{MPhaseGroundState}
\end{aligned}
\end{equation}
It is clear from these spin configurations that the \(\mathcal{M}\) phase is characterized by a non-vanishing staggered magnetization density given by
\begin{equation}
\begin{aligned} M_{\sigma} =  \frac{1}{N_c} \sum_{j = 1}^{N_c} (-1)^j \langle m_{\sigma,j} \rangle, \, \, m_{\sigma,j} \equiv \frac{1}{2}\big( \sigma_{j,A}^x -  \sigma_{j,B}^x  \big). 
\label{MPhaseStaggeredMagnetization}
\end{aligned}
\end{equation}

Fig. \ref{Alpha15CPhaseMagnetizations} shows that the \(\mathcal{C}\) phase is another ordered phase, which possesses a magnetic unit cell of 6 sites. We will see from our perturbation theory analysis that it is also characterized by a nonzero uniform chirality of the form:
\begin{equation}
\begin{aligned} \xi_{\sigma} 
\equiv  \frac{1}{N_c} \sum_{j = 1}^{N_c} \big\langle  \xi_{\sigma,j} \big\rangle, \, \xi_{\sigma,j} = 
\kappa_{\sigma,j} m_{\sigma,j + 1} - m_{\sigma,j} \kappa_{\sigma,j + 1},  
\\  \kappa_{\sigma,j} \equiv \frac{1}{2} \big( \sigma^y_{j,A} \sigma^y_{j,B}   + \sigma^z_{j,A} \sigma^z_{j,B} \big).
\label{CPhaseUniformChirality}
\end{aligned}
\end{equation}

From Fig. \ref{Alpha15BPhaseMagnetizations}, it is clear that restricting to odd sites only would reveal that \(\langle S_i^{x/y} \rangle\) alternate between the \(x\)- and \(y\)-directions with a single twist occurring in the middle of the chain, and this is characteristic of a topological soliton found in Ref. \cite{IsotropicKitaevUnderFieldPaper} for the isotropic chain. For PBCs, the \(\mathcal{B}\) phase is characterized by nonzero staggered chirality summed over all sites \(m\),
\begin{equation}
\begin{aligned} 
\gamma_{\sigma} \equiv \frac{1}{2 N_c} \sum_{m = 1}^{2 N_c}  (-1)^{m} \langle \gamma_{\sigma,m} \rangle, \,  \gamma_{\sigma,m} \equiv \big( \boldsymbol{\sigma}_{m} \times \boldsymbol{\sigma}_{m + 1} \big) \cdot \mathbf{\hat{z}}.
\label{BPhaseStaggeredChirality}
\end{aligned}
\end{equation}

In the PS state, the spins simply anti-align with the field as shown in Fig. \ref{Alpha15PSMagnetizations}. The correlation function \(C(r) = \langle S_1^x S_r^x \rangle\) has a characteristic period of 4, and due to the sublattice structure, the four-fold periodicity is associated with momentum \(q = \pi/a\). \(C(r)\) would show an exponential decay along every fourth \(r\) and a finite gap becomes evident \cite{IsotropicKitaevUnderFieldPaper}.  

\begin{figure}[!ht]
    \centering
        \centering   
       \begin{centering}     \includegraphics[width=0.86\columnwidth]{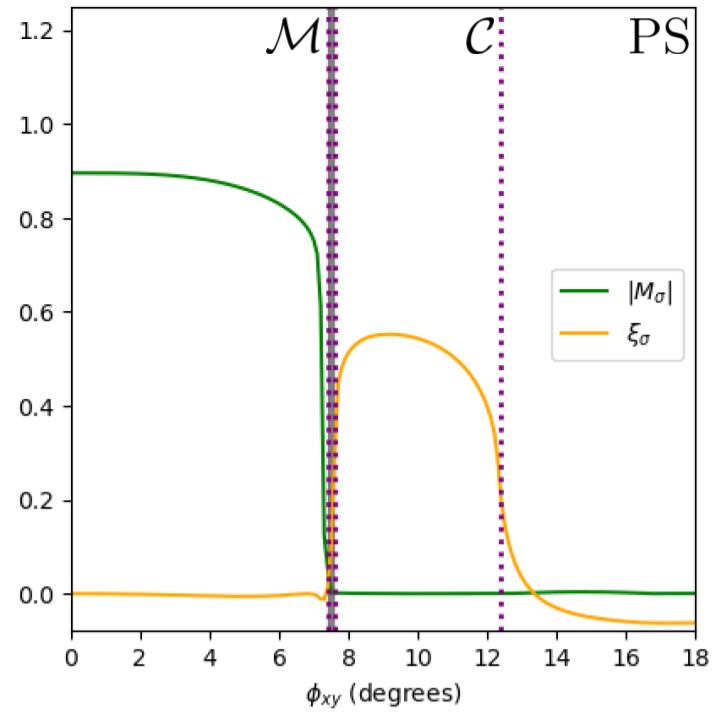}
       \end{centering}
  \caption{The magnitude of the staggered magnetization $|M_\sigma|$ and the uniform chirality $\xi_\sigma$ as a function of $\phi_{xy}$ for a fixed \(\alpha = 15^{\circ}\), \(h = 0.55\). Dashed purple lines mark the transition points between the \(\mathcal{M}\), \(\mathcal{C}\) and \(\mathrm{PS}\) phases, obtained with \(N = 200\) site DMRG. The tiny grey region in \(7.4 \leq \phi_{xy} \leq 7.6^{\circ}\) denotes a possibility of other phases which is beyond the current study.
  }
\label{ParametersPlot1}
\end{figure}

To understand the field angle dependence, we study the phase transition as a function of $\phi_{xy}$ for a given $h$ and $\alpha=15^\circ$. 
 For a fixed \(h = 0.55\) from \(0 \leq \phi_{xy} \leq 18^{\circ}\), which covers all of the \(\mathcal{M}\) (\(0 \leq \phi_{xy} \leq 7.4^{\circ}\)), \(\mathcal{C}\) (\(7.6 \leq \phi_{xy} \leq 12.4^{\circ}\)) and \(\mathrm{PS}\) (\(12.4 \leq \phi_{xy} \leq 18^{\circ}\))  phases,  
\(M_{\sigma}\) and $\xi_\sigma$ are finite in the \(\mathcal{M}\) and \(\mathcal{C}\) phases, respectively, as shown in Fig. \ref{ParametersPlot1}.
Dashed purple lines mark the transition points, and the grey region from \(7.4 \leq \phi_{xy} \leq 7.6^{\circ}\) denotes a region of possible incommensurate or disordered phases. In the \(\mathrm{PS}\), \(\xi_{\sigma}\) is small but finite, and is adiabatically connected to 0 at larger field directions. The transition between the \(\mathcal{C}\) phase and the \(\mathrm{PS}\) is then recognized by a sudden drop in \(\xi_\sigma\).

In Appendix \ref{MoreNumericalDataSection}, we show in Fig. \ref{CPhaseUnitCellUniformChiralities} the unit cell correlators \(\langle \xi_{\sigma,j} \rangle\) for various unit cells \(j\) in the middle of the chain for OBCS in the \(\mathcal{C}\) phase.
From this, it can be seen that the \(\mathcal{C}\) phase possesses nonzero staggered chirality as well (i.e. taking the sum of \(\langle  \xi_{\sigma,j} \rangle\) with a \((-1)^j\) factor yields a nonzero result), although this is a trivial effect attributed to the field. 

\begin{figure}[!ht]
    \centering
        \centering   
        \begin{centering} \includegraphics[width=0.86\columnwidth]{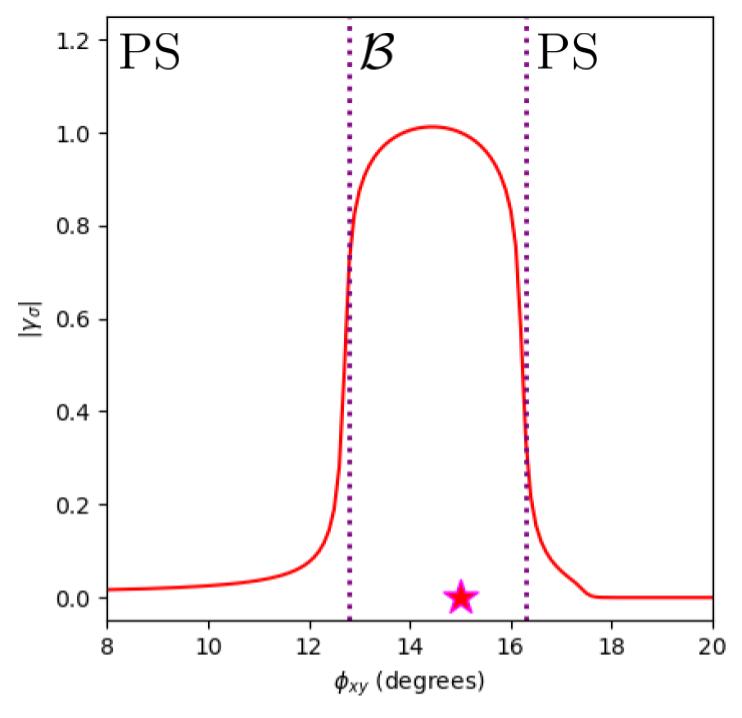}
        \end{centering} 
  \caption{ The magnitude of the staggered chirality $|\gamma_\sigma|$ as a function of $\phi_{xy}$ for a fixed \(\alpha = 15^{\circ}\), \(h = K/2 \sim 0.707\).  Dashed purple lines mark the transition points between the \(\mathcal{B}\) and \(\mathrm{PS}\) phases, obtained with \(N = 200\) site DMRG, and the pink star marks the exactly solvable point for the \(\mathcal{B}\) phase discussed in Sec. \ref{exactlysolvablepointsection}.}
\label{ParametersPlot2}
\end{figure}

Fig. \ref{ParametersPlot2} shows the transition between the \(\mathcal{B}\) and \(\mathrm{PS}\) phases for the field strength \(h = K/2\) from \(8^{\circ} \leq \phi_{xy} \leq 20^{\circ}\), which covers the \(\mathcal{B}\) phase (\(12.8^{\circ} \leq \phi_{xy} \leq 16.3^{\circ}\)) and \(\mathrm{PS}\) (\(8^{\circ} \leq \phi_{xy} \leq 12.8^{\circ}\), \(12.8^{\circ} \leq \phi_{xy} \leq 20^{\circ}\)) phases. As shown, $\gamma_\sigma$ is finite in the \(\mathcal{B}\) phase. Dashed purple lines mark the transition points and the pink star at \(\phi_{xy} = \alpha\) marks the exactly solvable point inside the \(\mathcal{B}\) phase, which we will discuss later in Sec. \ref{exactlysolvablepointsection}. 

\vfill

\section{Understanding Ordered Phases - Perturbation Theory\label{perturbationtheorysection}}
In this section, we aim to achieve an analytical understanding of the two ordered phases \(\mathcal{M}\) and \(\mathcal{C}\). To do so, we employ a perturbation theory where we derive a low-energy effective model in the strong anisotropic regime. For convenience, we align the spin quantization axis along the \(x\)-direction, i.e. in the direction of the strong \(K_x\) bonds. \\
\indent We first note that when \(K_y = 0\), i.e. the unit cells are disconnected, the model given by Eq. \eqref{FullHamiltonian} is exactly solvable under a field applied in the pure \(x\)-direction, i.e. \(\mathbf{h} = h_x \hat{\mathbf{x}}\). In this case, the Hamiltonian possesses the eigenstates \((|m_{j,A} \rangle \otimes | m_{j,B} \rangle )^{\otimes N_c}\) with \(m \in \{ \uparrow,  \downarrow \} \). When \(h_x < K_x/2\), the ground state is \(2^{N_c}\)-fold degenerate with energy \(E_0 = -N_c K_x/4\) and wavefunctions  \(| \Psi_0 \rangle = (| \psi_j \rangle)^{\otimes N_c}\)
with \( | \psi_j \rangle \in \{ | a_j \rangle, | b_j \rangle \}\), where
\begin{equation}
\begin{aligned}
    |a_j \rangle \equiv  | \uparrow \downarrow \rangle_j, \, |b_j \rangle \equiv | \downarrow \uparrow \rangle_j.
        \label{EffectiveSpins}
\end{aligned}
\end{equation}
When \(h_x > K_x/2\), the ground state is unique, and all of the unit cells polarize against the field. The ground state energy in this case is \(E_0 = N_c ( K_x/4 - h_x ) \) with wavefunction  \(| \Psi_0 \rangle = (|\downarrow \downarrow \rangle)^{\otimes N_c}\).
At the special point \(h_x = K_x/2\), the ground state is \(3^{N_c}\)-fold degenerate, but carries additional \(U(1)\) symmetry in that the direction of the spin is arbitrary. \\
\indent We proceed to perform a perturbation theory for intermediate \(h_x\) where we take \(K_x/2 > h_x \gg K_y, h_y\).
In this case, we split the Hamiltonian as \(\mathcal{H} = \mathcal{H}_0 + \mathcal{H}_1\) where
\begin{equation}
\begin{aligned}
    & \mathcal{H}_0 \equiv \sum_{j=1}^{N_c} \left[ \frac{K_x}{4} \sigma_{j\A}^x \sigma_{j\B}^x + \frac{h_x}{2} (\sigma_{j\A}^x + \sigma_{j\B}^x)\right], \\ & \mathcal{H}_1  \equiv \sum_{j=1}^{N_c} \left[  \frac{K_y}{4} \sigma_{j\B}^y \sigma_{j+1,\A}^y  + \frac{h_y}{2} (\sigma_{j\A}^y + \sigma_{j\B}^y)\right].  
    \label{H0H1Definitions}
\end{aligned}
\end{equation}
We wish to obtain a third-order low-energy effective Hamiltonian \(\boldsymbol{\mathcal{H}}_{\mathrm{eff}} = \sum_{n = 0}^{3} \mathcal{H}_{\mathrm{eff}}^{(n)} \) containing the terms 
\begin{equation}
\begin{aligned}
    \mathcal{H}_{\mathrm{eff}}^{(n)} = \begin{Bmatrix} \mathcal{P} \mathcal{H}_0 \mathcal{P} = E_0 \mathcal{P}, \, n = 0 \\ \mathcal{P}\bigg[ \mathcal{H}_1 \mathcal{Q} \frac{1}{E_0 - \mathcal{H}_0} \mathcal{Q} \bigg]^{n - 1} \mathcal{H}_1 \mathcal{P}, \, n = 1, 2, 3       
    \end{Bmatrix},   
    \label{EffectiveHamiltonianOperator}
\end{aligned}
\end{equation}
where \(\mathcal{P}\) projects onto the low-energy subspace, and \(\mathcal{Q} = 1 - \mathcal{P}\). We note that \(|a_j \rangle\) and \(|b_j \rangle\) defined in Eq. \eqref{EffectiveSpins} may be taken to be effective up and down spins in the low-energy subspace, and thus we may write the result for \(\mathcal{H}_{\mathrm{eff}}\) in terms of Pauli matrices defined in terms of these effective spins:
\begin{equation}
\begin{aligned}
    \tau_j^x &\equiv |a_j \rangle  \langle b_j | +  |b_j \rangle  \langle a_j |, \\ 
    \tau_j^y & \equiv -i |a_j \rangle  \langle b_j | + i|b_j \rangle  \langle a_j |, \\ 
    \tau_j^z & \equiv |a_j \rangle  \langle a_j | - |b_j \rangle  \langle b_j |. \label{EffectivePauliMatrices}
\end{aligned}
\end{equation}

Due to the presence of the field, at third order, it is enough to consider a cluster of two dimers connected by a single \(y\)-bond, and later add their contributions. Contributions due to a cluster of three or more dimers with two or more \(y\)-bonds do not arise until higher order in this expansion. With the quantization axis along the \(x\)-axis, the \(K_y\) term of \(\mathcal{H}_1\) flips the inner spins of two neighboring dimers while each of the \(h_y\) terms only flips one of the four. At the end of the perturbative processes, we must return to one of the ground states in order to survive the final projection \(\mathcal{P}\) and generate terms of \(\mathcal{H}_{\mathrm{eff}}\).  At second order, non-trivial terms are obtained by either two  applications of the \(K_y\) term on the neighboring inner spins, or two applications of the \(h_y\) terms on the cluster. At third order, non-trivial terms are obtained by the combination of the application of the \(K_y\) term on the neighbouring inner spins, and two applications the \(h_y\) terms on the cluster. 

Carrying out these processes, the resulting third-order effective Hamiltonian is found to be
\begin{equation}
\begin{aligned} \boldsymbol{\mathcal{H}}_{\mathrm{eff}}[\mathbf{h}]  = \sum_{j = 1}^{N_c} \Big[    \tilde{K}_T  \tau^z_{j} \tau^z_{j + 1}  + \tilde{K}_L \tau^x_{j} \tau^x_{j + 1}           \\ +\tilde{K}_{DM} \big(\tau^x_{j} \tau^z_{j + 1} - \tau^z_{j} \tau^x_{j + 1} \big) - \tilde{H}    \tau^x_{j}   \Big],
\label{ThirdOrderEffectiveHamiltonian}
\end{aligned}
\end{equation}
with constants defined to be
\begin{equation}
\begin{aligned}   &\tilde{K}_T \equiv   \bigg( \frac{h_x^2 }{8  K_x( K_x^2 - 4h_x^2)}  \bigg) K_y^2 -  \bigg( \frac{4 h_x^2}{( K_x^2 - 4h_x^2)^2 } \bigg)   K_y h_y^2, \\ &\tilde{K}_L \equiv \bigg( \frac{K_x^2  }{(K_x^2 - 4 h_x^2)^2} \bigg) K_y h_y^2, \\ &\tilde{K}_{DM} \equiv \bigg( \frac{2 K_x  h_x }{(K_x^2 - 4 h_x^2)^2} \bigg) K_y h_y^2,  \\
&\tilde{H} \equiv   \bigg(  \frac{2 K_x}{K_x^2 - 4h_x^2}\bigg) h_y^2  -   \bigg(  \frac{2 K_x^2}{(K_x^2 - 4h_x^2)^2}\bigg)K_y h_y^2. 
\label{ThirdOrderEffectiveHamiltonianConstants}
\end{aligned}
\end{equation}
At zero field \(\mathbf{h} = \mathbf{0}\), \(\tilde{K}_T\), \(\tilde{K}_L\), \(\tilde{K}_{DM}\) and \(\tilde{H}\) all vanish, and  \(\boldsymbol{\mathcal{H}}_{\mathrm{eff}}\) is simply proportional to the identity. In fact, it can be shown that it is proportional to the identity at all orders of perturbation theory. This shows that the \(2^{N_c}\)-fold degeneracy found from the vanishing of two of the Majorana fermion energy bands given by Eq. \eqref{MajoranaFermionsEnergyBands}
cannot be lifted in this case, and this is due to the fact that is protected by non-local SU(2) symmetries \cite{ZeroFieldChainPaper,sen2010spin}.   

\subsection{Understanding the ordered phase $\mathcal{M}$ with effective staggered magnetization}
As noted from the DMRG results, the ordered phase denoted by $\mathcal{M}$ is characterized by a non-vanishing staggered magnetization density between the two sublattices, Eq. \eqref{MPhaseStaggeredMagnetization}. 
In terms of the effective spin \(\tau\), the phase exhibits a staggered magnetization given by,
\begin{equation}
\begin{aligned} M_{\tau} \equiv \frac{1}{N_c}\sum_{j = 1}^{N_c}  (-1)^j \langle  \tau_j^z \rangle. 
\label{EffectiveStaggeredMagnetization}
\end{aligned}
\end{equation}
To understand the emergence of this ordered phase, we first investigate the simplest limit of $h_y = 0$ below.

\subsubsection{Low field along \(h_y = 0\) phase space}
\indent 
If we set \(h_y = 0\), we simply find from the effective \(\tau\)-model \(\boldsymbol{\mathcal{H}}_{\mathrm{eff}}\) the antiferromagnetic Ising model,
\begin{equation}
\begin{aligned} \boldsymbol{\mathcal{H}}_{\mathrm{eff}}[\mathbf{h} = h_x \mathbf{\hat{x}}]  =  \sum_{j = 1}^{N_c} \Big[ \tilde{K}_T  \tau^z_{j} \tau^z_{j + 1}     \Big].
\label{ThirdOrderEffectiveHamiltonianhyEqual0}
\end{aligned}
\end{equation}
The ground state in this case is two-fold degenerate, with the effective spins displaying Néel order. It is thus immediately clear that the \(\mathcal{M}\) phase is characterized by non-vanishing staggered magnetization density \(M_{\tau}\).  
In writing the ground state in terms of original spins, we find the spin configurations given by Eq. \eqref{MPhaseGroundState}.    

\subsubsection{Classical analysis along \(h_y = 0\) line - vanishing of \(\mathcal{M}\) phase in isotropic limit}
Evidently, from the \(\alpha = 45^{\circ}\) isotropic Kitaev chain phase diagram \cite{Sorensen2023PRRa}, the $\mathcal{M}$ phase vanishes. In the isotropic limit, the system immediately polarizes under low fields applied along any direction except  $\phi_{xy} = 45^\circ$.  
Furthermore,  for any \(\alpha \neq 45^{\circ}\), the \(\mathcal{M}\) phase would occupy some region of the \(h_y = 0\) line as shown in Fig. \ref{Alpha15AllMagnetizations} with $\alpha = 15^\circ$. To understand the dependence on  $\alpha$, we study the classical phase diagram for all $\alpha \leq 45^{\circ}$ with a limited phase space restricted to the $h_y=0$ line. 

For small finite \(h_x\) along the \(h_y = 0\) line, we assume that the classical ground state is the classical analog of the ordered phase \(\mathcal{M}\) spin configuration given by Eq. \eqref{MPhaseGroundState}, i.e. 
\begin{equation}
\begin{aligned} 
\mathbf{S}_{2j-1,A} &= \pm S \mathbf{\hat{x}}, \,  \mathbf{S}_{2j-1,B} = \mp S \mathbf{\hat{x}}, \\  \mathbf{S}_{2j,A}  &=  \mp S \mathbf{\hat{x}}, \,  \mathbf{S}_{2j,B} = \pm S \mathbf{\hat{x}}, \;\;\;\; 
j = 1, ..., N_c/2, 
\label{ClassicalMPhaseSpinConfiguration}
\end{aligned}
\end{equation}
where \(S\) is the magnitude of the spin. We employ a Holstein-Primakoff transformation and perform a linear spin wave expansion around these classical states in order to derive the magnon branches in the classical \(\mathcal{M}\) phase at \(h_y = 0\). Details of the calculation are given in Appendix \ref{linearspinwavetheoryappsection}. The four resulting branches are
\begin{equation}    
\begin{aligned}
\omega_{(+),\pm} =   \sqrt{\bigg( \frac{h_x}{S} + K_x\bigg) \bigg( \frac{h_x}{S} + K_x \pm K_y \bigg)}, \\ \omega_{(-),\pm} =   \sqrt{\bigg( \frac{h_x}{S} - K_x\bigg) \bigg( \frac{h_x}{S} - K_x \pm K_y \bigg)}.     
\label{MPhaseMagnonBranches}
\end{aligned}
\end{equation}
The lowest value of the field \(h_x > 0\) for which one of the magnon branches vanishes is \(h_x = S(K_x - K_y)\) (in particular, \(\omega_{(-),+}\) vanishes). Thus, when \(h_y = 0\), the boundary where the  \(\mathcal{M}\) phase becomes unstable is given by 
\begin{equation}    
\begin{aligned}
h_{x,c_1} = S(K_x - K_y), \, K_x \geq K_y,     
\label{ClassicalMPhaseBoundary}
\end{aligned}
\end{equation} 
confirming that the \(\mathcal{M}\) phase disappears when \(K_x = K_y\). 

\begin{figure}[!ht]
    \centering
        \centering
        \begin{centering}        \includegraphics[width=0.91\columnwidth]{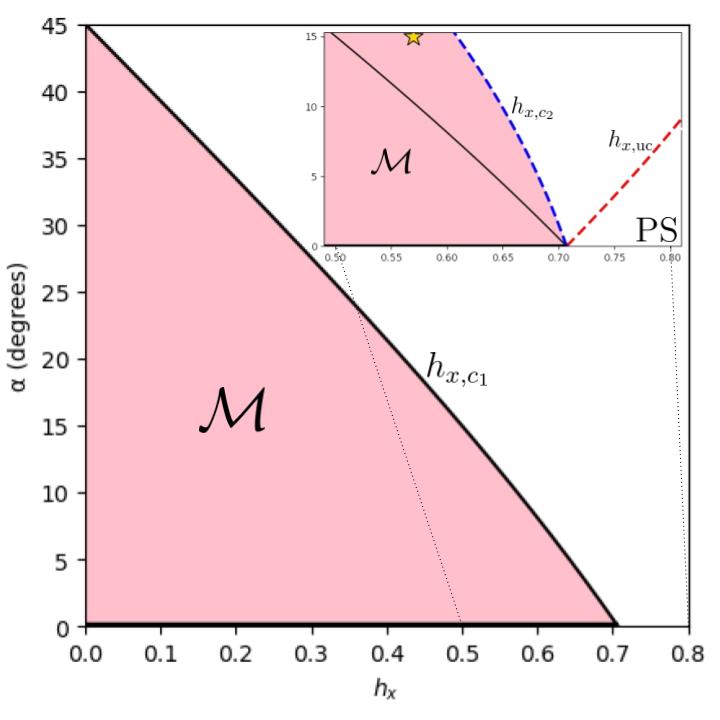}
        \end{centering}
  \caption{The phase diagram with \(h_y = 0\) as a function of $h_x$ and $\alpha$. The phase boundary \(h_{x,c_1}\)  
  above which the \(\mathcal{M}\) phase is unstable obtained by the classical spin wave analysis is shown by the solid black line. The black horizontal line at \(\alpha = 0^{\circ}\) indicates the limit where the theory does not apply as $K_y=0$.
  The inset is a zoom-in of the region 
  $0.49 \leq h_x \leq 0.81$. The dashed blue line is the phase boundary \(h_{x,c_2}\) obtained from the fourth order effective \(\tau\)-model within the small \(\alpha\) regime, which indicates that the $\mathcal{M}$ phase is extended beyond the classical regime. The red dashed line is the classical phase boundary \(h_{x,\mathrm{uc}}\) below which the PS is unstable. 
  The gold star indicates a LU phase for a fixed $\alpha = 15^{\circ}$ and $h_x = 0.57$ obtained from DMRG.}
\label{hxEqual0ClassicalPhaseBoundary}
\end{figure}

The critical field \(h_{x,c_1}\) above which the \(\mathcal{M}\) phase becomes unstable in the \((h_x, \alpha)\) phase space is shown in Fig. \ref{hxEqual0ClassicalPhaseBoundary}. We take \(S = 1/2\) for comparison to the \(S = 1/2\) quantum phase diagram. The black line extending from \(h_x = 0\) to \(h_x = 
K_x/2\) when \(\alpha = 0^{\circ}\) indicates that there is no \(\mathcal{M}\) phase along this line, as the classical ground state maintains the \(2^{N_c}\)-fold degeneracy of the quantum ground state. 
For \(\alpha \neq 0^{\circ}\), classically, it appears that the Hamiltonian is also \(2^{N_c}\)-fold degenerate, with any configuration \(S_{j,A/B} = + \eta_j S \mathbf{\hat{x}}, \, S_{j,B} = -\eta_j S \mathbf{\hat{x}}, \, \eta_j = \pm 1\)  yielding the same minimum classical energy \(E_{\mathrm{cl}} = -N_c S^2 K_x\). It is the order by disorder phenomena that selects the two configurations given by Eq. \eqref{ClassicalMPhaseSpinConfiguration}, due to quantum fluctuations as derived in Appendix \ref{linearspinwavetheoryappsection}.  

The dashed blue line in the inset of Fig. \ref{hxEqual0ClassicalPhaseBoundary} is another estimate of the critical field  \(h_{x,c_2}\) above which the \(\mathcal{M}\) phase becomes unstable, valid only in the strong anisotropic regime, i.e., small $\alpha$ region. This quantity is obtained from a fourth order effective \(\tau\)-model \(\boldsymbol{\mathcal{H}}_{\mathrm{eff}}^{\mathrm{4}^{\mathrm{th}}}\) with \(h_y = 0\) that we will discuss in Sec. \ref{luphasediscussion}. We will show that 
the \(\mathcal{M}\) phase transitions to a \(\mathrm{LU}\) phase that possesses a magnetic unit cell of 8 sites. The gold star indicates a point of a \(\mathrm{LU}\) phase that possesses a magnetic unit cell of 10 sites, as observed from DMRG, indicating that the phase space of the ${\mathcal M}$ phase is overestimated in the perturbation theory most likely due to missing higher order terms.

We also investigate the critical field below which the PS is unstable. Starting from the PS at high field, where
all of the spins are polarized due to the external field, i.e.
\begin{equation}
\begin{aligned} \mathbf{\hat{S}}_{j,A} = \mathbf{\hat{S}}_{j,B} = - S \mathbf{\hat{h}}, \, j = 1, ..., N_c,
\label{ClassicalAntialignedState}
\end{aligned}
\end{equation}
we found that the magnon branches of the PS are identical to 
the magnon branches \(\omega_{(-),\pm}\) of the \(\mathcal{M}\) phase when \(h_y = 0\). 
The critical field below which the PS becomes unstable is given by
\begin{equation}    
\begin{aligned}
h_{x,\mathrm{uc}} = S(K_x + K_y).      
\label{AntialignedStateClassicalPhaseBoundary}
\end{aligned}
\end{equation}
The critical field $h_{x,\mathrm{uc}}$ is shown by the dashed red line in the inset of Fig. \ref{hxEqual0ClassicalPhaseBoundary}.
The details of the derivation is presented in Appendix \ref{linearspinwavetheoryappsection}.
This analysis suggests that other phases such as the \(\mathrm{LU}\) phases may occur in between \(h_{x,c_1}\) and \(h_{x,\mathrm{uc}}\). However, as $\alpha$ increases, the quantum fluctations may affect these intermediate phases, and it is known that at $\alpha = 45^\circ$, the transition to the PS occurs as soon as the field is finite \cite{Sorensen2023PRRl}. Conducting further classical calculations as to the precise whereabouts of these phases is left for future work.  

\subsubsection{Low nonzero fields \((h_x, h_y)\) regime}
\indent At low field strengths \(h\), for some range of small \(\phi_{xy}\), we take \(h_x \ll K_x/2\) and we would find that \(\tilde{K}_{DM}, \tilde{K}_{L} \ll \tilde{H}, \tilde{K}_{T}\). In this limit, we would find that the effective \(\tau\)-model \(\mathcal{H}_{\mathrm{eff}}\) 
reduces to the antiferromagnetic transverse field Ising model,
\\
\begin{equation}
\begin{aligned} \boldsymbol{\mathcal{H}}_{\mathrm{eff}}[\mathbf{h}]  \simeq \sum_{j = 1}^{N_c} \bigg[  \tilde{K}_T  \tau^z_{j} \tau^z_{j + 1} - \tilde{H}    \tau^x_{j}  \bigg], \, \, h_x \ll \frac{K_x}{2}.
\label{ThirdOrderEffectiveHamiltonianLowFieldLimit}
\end{aligned}
\end{equation}

The exact spectrum of \(\boldsymbol{\mathcal{H}}_{\mathrm{eff}}\) in this limit can be obtained through Jordan-Wigner transformation and Majorana fermions \cite{pfeuty1970one}. The result is found to be 
\begin{equation}
\begin{aligned} \varepsilon (k) = 2 \sqrt{\tilde{K}_{T}^2 + \tilde{H}^2 + 2 \tilde{K}_{T} \tilde{H} \cos(k) }  , \, \, h_x \ll \frac{K_x}{2}.
\label{LowFieldLimitExactSpectrum}
\end{aligned}
\end{equation}
The resulting gap \(\Delta\) between the ground state and the first excited state occurs at \(k = \pi\), and we would find \(\Delta = 2 | \tilde{K}_{T} - \tilde{H}|\). 
A second order quantum phase transition in this case is found to occur when \(\tilde{H} = \tilde{K}_T\), where \(\Delta\) vanishes  \cite{fradkin2013field, QPTsBook, chakrabarti2008quantum}. By Eq. \eqref{ThirdOrderEffectiveHamiltonianConstants}, this yields the critical field direction 
\begin{equation}
\begin{aligned} \phi_{xy,c} \simeq \arctan\bigg( \frac{K_y}{4\sqrt{K_x(K_x - K_y)}} \bigg).
\label{LowFieldCriticalAngle}
\end{aligned}
\end{equation}
For small fixed \(h\), even in the limit as \(h \rightarrow 0\), \(\phi_{xy,c} \) gives a phase boundary separating the \(\mathcal{M}\) phase and the \(\tau\)-polarized state across \(\phi_{xy,c} \).

\subsection{Understanding the ordered phase $\mathcal{C}$ with effective uniform chirality}
In the strong anisotropic regime, the ordered chiral phase $\mathcal{C}$ appears at higher \(h_x\) with a small \(h_y\) component. Here, all nontrivial terms of \(\boldsymbol{\mathcal{H}}_{\mathrm{eff}}\) - the effective field (${\tilde H}$), transverse Ising interaction (${\tilde K}_T$), longitudinal Ising interaction (${\tilde K}_L$) and DM interaction (${\tilde K}_{DM}$) 
- all become prominent. To gain insight, we begin with a simpler model by setting ${\tilde H}=0$, i.e., the $h_y=0$ case, and explore how the chiral phase $\mathcal{C}$ emerges. We then proceed to show that it persists when $h_y \neq 0$.  

We first note that the effective \(\tau\)-model with ${\tilde H}=0$ is exactly solvable.
It maps to a single component Fermi gas of spinless fermions on a 1D lattice via a Jordan-Wigner transformation \cite{XYModelDMInteractionPaper, barouch1970statistical, barouch1971statistical, lieb1961two}. When \(\tilde{H} = 0\), the effective Hamiltonian \(\boldsymbol{\mathcal{H}}_{\mathrm{eff}}\) can be written as  
\begin{equation}
\begin{aligned} 
\boldsymbol{\mathcal{H}}_{\mathrm{eff}}[\mathbf{h}] \bigg|_{\tilde{H} = 0}  = & \frac{J}{4}\sum_{j = 1}^{N_c} \Big\{  (1 - \gamma)  \tau^z_{j} \tau^z_{j + 1} \ + (1 + \gamma) \tau^x_{j} \tau^x_{j + 1}  \\ 
& + d \big(\tau^x_{j} \tau^z_{j + 1} - \tau^z_{j} \tau^x_{j + 1} \big)     \Big\},
\label{ThirdOrderEffectiveHamiltonianNoH}
\end{aligned}
\end{equation}
where the exchange interactions are defined to be 
\begin{equation}
\begin{aligned} J \equiv 2(\tilde{K}_L + \tilde{K}_T), \, \gamma \equiv \frac{\tilde{K}_L - \tilde{K}_T}{\tilde{K}_L + \tilde{K}_T}, \, d \equiv  \frac{2 \tilde{K}_{DM}}{\tilde{K}_L + \tilde{K}_T}.
\label{NewEffectiveExchangeInteractions}
\end{aligned}
\end{equation}

\begin{figure}[!ht]
    \centering
        \centering 
         \begin{centering} \includegraphics[width=0.88\columnwidth]{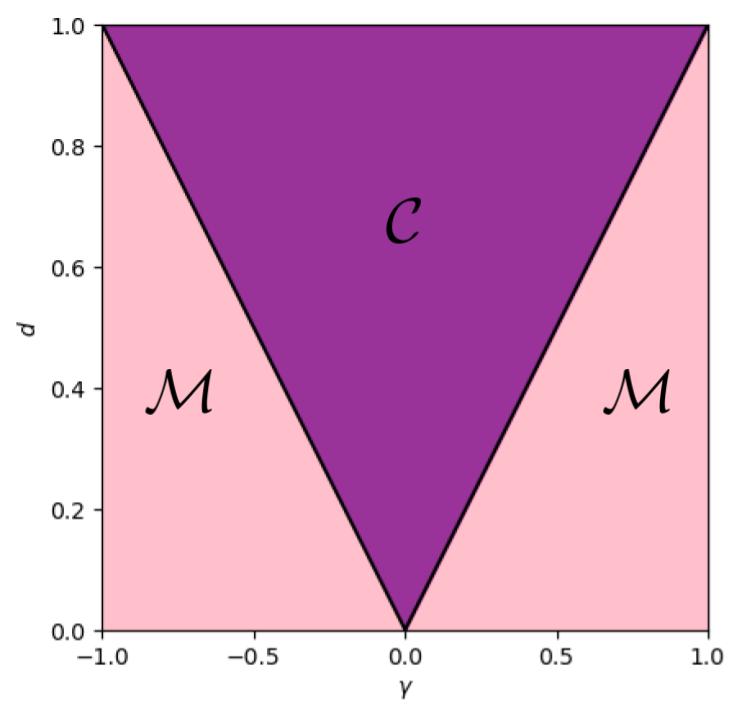}
         \end{centering}
  \caption{Phase diagram of the effective \(\tau\)-model under zero effective field \(\tilde{H} = 0\), revealing the \(\tau\)-chiral \(\mathcal{C}\) and the \(\tau\)-antiferromagnetic phases \(\mathcal{M}\). }
\label{EMPhaseDiagramZeroH}
\end{figure}

Upon carrying out the Jordan-Wigner transformation on \(\boldsymbol{\mathcal{H}}_{\mathrm{eff}}\) when \(\tilde{H} = 0\),
the single particle excitation spectrum is found to be
\begin{equation}
\begin{aligned} \varepsilon(k) = J \bigg( d \sin (k) + \sqrt{\cos^2 (k) + \gamma^2 \sin^2 (k)} \bigg).
\label{NoHExactSpectrum}
\end{aligned}
\end{equation}
For \(0 \leq d < |\gamma| \), \(\varepsilon(k)\) is positive for all \(k\) and the ground state of the model is the vacuum of the corresponding Bogoliubov operators, which is independent of the value of \(d\) \cite{XYModelDMInteractionPaper, derzhko2006dynamic}. The ground state energy, and hence all two-site correlators and the magnetization of the zero-temperature state of the model are thus insensitive to the DM interaction in this case. However, for \(d > |\gamma|\) (i.e. \(\tilde{K}_{\mathrm{DM}} > \frac{1}{2} \big| \tilde{K}_{\mathrm{L}} - \tilde{K}_{\mathrm{T}} \big| \) ), \(\varepsilon(k)\) is found to go negative within the range \(|k + \pi/2| <  \arccos \big[ (1 + d^2 - \gamma^2)^{-1/2} \big] \), and so modes in this case have to be filled to construct the ground state, which now depends on \(d\). The phase diagram in \((\gamma, d)\) phase space is shown in Fig. \ref{EMPhaseDiagramZeroH}. It is important to note that the quantity given by 
\begin{equation}
\begin{aligned} \xi_{\tau} 
\equiv \frac{1}{N_c} \sum_{j = 1}^{N_c} \big\langle \big( \boldsymbol{\tau}_{j} \times \boldsymbol{\tau}_{j + 1} \big) \cdot \mathbf{\hat{y}} \big\rangle   
\label{EffectiveUniformChirality}
\end{aligned}
\end{equation}
is identical to the uniform chirality quantity Eq. \eqref{CPhaseUniformChirality} found in the \(\mathcal{C}\) phase. The chirality \(\xi_{\tau}\) is evaluated analytically and found to be 
\begin{equation}
\begin{aligned} \xi_{\tau}
= \begin{Bmatrix}
    0, \, \, d \leq |\gamma| \\ \frac{4}{\pi} \sqrt{\frac{d^2 - \gamma^2}{1 + d^2 - \gamma^2}} , \, \, d >  |\gamma|
\end{Bmatrix}.
\label{EffectiveUniformChiralityNoHEvaluated}
\end{aligned}
\end{equation}

Furthermore, DMRG studies with a small added alternating field term \(\sum_{j = 1}^{N_c} (-1)^j h_z \tau_j^z\) reveals that \(M_{\tau} \neq 0 \) when \(d \leq  |\gamma|\) and \(M_{\tau} = 0 \) when \(d > |\gamma|\). \(d = |\gamma|\) is therefore a critical line which separates a \(\tau\)-chiral phase \(\mathcal{C}\) characterized by nonzero \(\xi_{\tau}\) and zero \(M_{\tau}\) when \(d >  |\gamma|\), and the \(\tau\)-antiferromagnetic phase \(\mathcal{M}\) characterized by zero \(\xi_{\tau}\) and nonzero \(M_{\tau}\) when \(d <  |\gamma|\), as shown in Fig. \ref{EMPhaseDiagramZeroH} \cite{XYModelDMInteractionPaper, tian2018quantum, zhong2013effects}.

We note that when \(\tilde{H} = 0\), the \(\tau\)-model is gapless in the \(\mathcal{C}\) phase.
This is because the spectrum \(\varepsilon (k)\) vanishes when \(k = - \pi/2 \pm \arccos \big[ (1 + d^2 - \gamma^2)^{-1/2} \big] \) for all \(d > |\gamma|\). However, nonzero \(\tilde{H}\) in addition to higher order terms introduces a gap in the spectrum of our model. 

We now turn on a finite $h_y$ leading to a finite ${\tilde H}$, and explore the persistence of the ${\mathcal C}$ phase.
To do so, we perform finite DMRG on the full effective \(\tau\)-model \(\boldsymbol{{\mathcal{H}}}_{\mathrm{eff}}\).
We take \(N_c = 100\) sites and select intermediate field strengths \(h\) and small field angles  \(\phi_{xy}\) within the small \(\alpha\) regime where the perturbation theory remains valid and higher order terms do not come heavily into play. 

\begin{figure}[!ht]
    \centering
    \captionsetup[subfigure]{justification=centering}
    \begin{subfigure}[b]{0.4\textwidth}        \centering        
        \caption{\(h = 0.01\)}
        \begin{centering}        \includegraphics[width=1\textwidth]{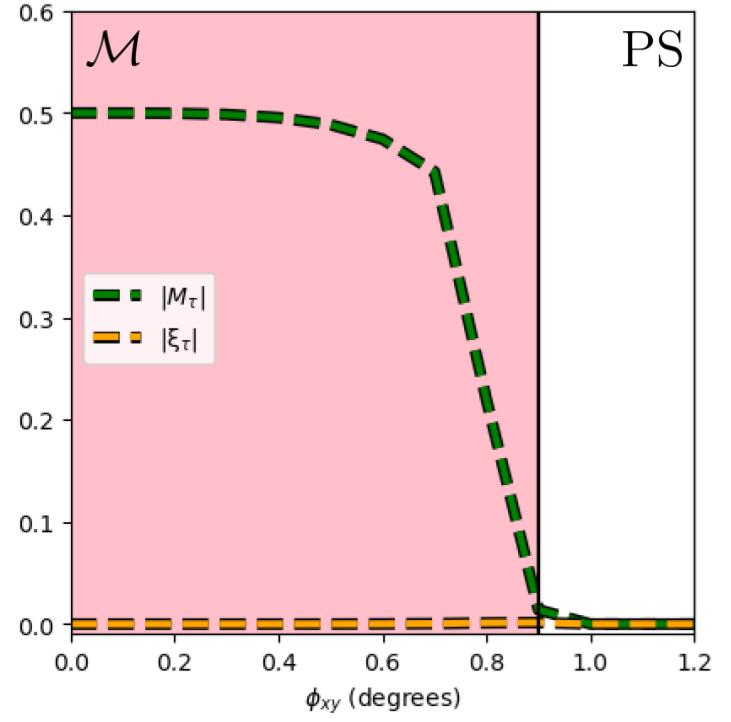}
        \end{centering}
    \label{EMParametersLowField}
     \end{subfigure}
     \begin{subfigure}[b]{0.4\textwidth}
         \centering       
         \caption{\(h = 0.66\)}
            \begin{centering}            \includegraphics[width=1\textwidth]{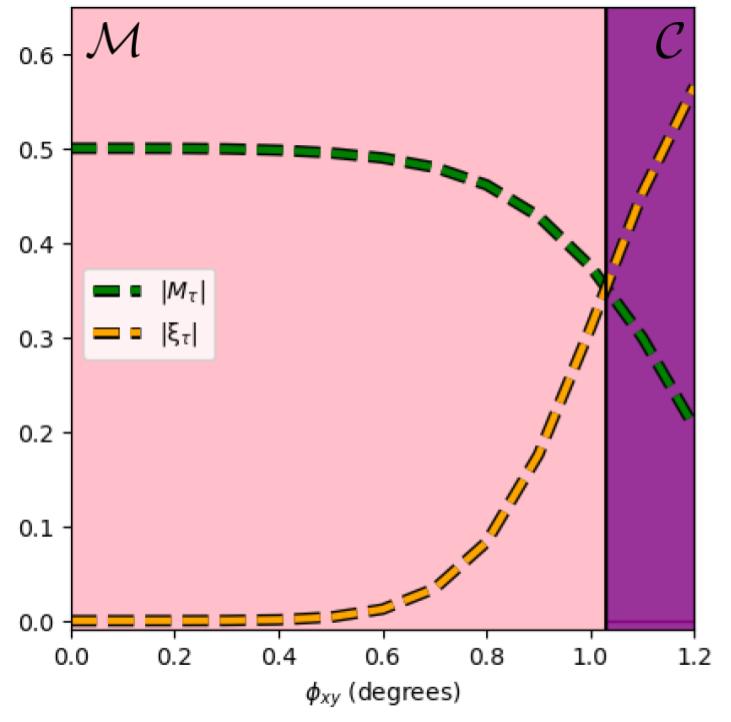}  
            \end{centering}
         \label{EMParametersHigherField}
     \end{subfigure}
  \caption{ The magnitude of the staggered magnetization \(M_{\tau}\) and uniform chirality \(\xi_{\tau}\) of the effective \(\tau\)-model as a function of \(\phi_{xy}\)  
  obtained with \(N_c = 100\) site DMRG at fixed low and intermediate field strengths with \(\alpha = 3^{\circ}\). Black vertical lines separating colored panels mark the predicted transition of the effective \(\tau\)-model.} 
\label{EMParametersPlot}
\end{figure}

To observe the persistence of the \(\mathcal{C}\) phase, we examine the magnitude of the quantities \(M_{\tau}\) and \(\xi_{\tau}\). Fig. \ref{EMParametersPlot} shows \(|M_{\tau}|\) (green dashed line) and \(|\xi_{\tau}|\) (orange dashed line) at (a) low \((h = 0.01)\) and (b) higher \((h = 0.66)\) field strengths for a fixed small \(\alpha = 3^{\circ}\) (which we choose so that \(K_y\) is approximately \(5\%\) of \(K_x\)) from \(0 < \phi_{xy} < 1.15^{\circ}\). 

At \(h = 0.01\), the black line at \(\phi_{xy} = 0.9^{\circ}\) marks the \(\tau\)-model transition beyond which the \(\mathcal{M}\) phase vanishes. This transition point approximately agrees with \(\phi_{xy,c}\) given by Eq. \eqref{LowFieldCriticalAngle} (\(\phi_{xy,c} \simeq 0.77^{\circ}\) here), obtained from the low fields approximation of the \(\tau\)-model given by Eq. \eqref{ThirdOrderEffectiveHamiltonianLowFieldLimit}. It is the inclusion of the third order terms that improves the estimate. The chirality \(\xi_{\tau} = 0\) throughout this phase space. 

At \(h = 0.66\), the black line at \(\phi_{xy} \simeq 1.03^{\circ}\) marks the \(\tau\)-model transition from the \(\mathcal{M}\) phase to the \(\mathcal{C}\) phase, where \(M_{\tau} \) approaches zero and \(\xi_{\tau}\) now approaches a nonzero value. Due to the neglected higher order terms, the \(\tau\)-model transition does not align with the transition of the original \(\sigma\)-model. Despite such a quantitative difference, the \(\tau\)-model captures the ${\mathcal C}$ phase with uniform chirality, which originates from the DM interaction
${\tilde K}_{DM}$. This is consistent with previous findings that the DM interaction gives rise to magnetic chirality in quantum spin systems \cite{XYModelDMInteractionPaper, soltani2019ising, guo2023spin, camley2023consequences, chen2021observation}. 

\section{Understanding Phase \(\mathcal{B}\) with soliton - Exactly Solvable Points \label{exactlysolvablepointsection}}
The Hamiltonian given by Eq. \eqref{FullHamiltonian} with \(h_z = 0\) can be solved exactly at several special values of the field \((h_x, h_y)\). Namely, at the points
\begin{equation}
\begin{aligned} \big(h_x^*, h_y^* \big) = \bigg(\frac{K_x}{2},  \frac{K_y}{2}\bigg) \, \leftrightarrow \, \big(h^*, \phi_{xy}^* \big) = \bigg(\frac{K}{2}, \alpha\bigg),
\label{HigherFieldEsps}
\end{aligned}
\end{equation}
under PBCs, \(\mathcal{H}\) may be written as 
\begin{equation}
\begin{aligned}
\mathcal{H}^* =   &\frac{1}{4} \sum_{j=1}^{N_c} \bigg[ K_x \big(\sigma_{j,\A}^x  + 1 \big) \big( \sigma_{j,\B}^x + 1 \big) \\
&    + K_y \big( \sigma_{j,\B}^y + 1\big) \big(  \sigma_{j+1,\A}^y  + 1 \big) \bigg]  \\
  &  \;\; - \frac{N_c (K_x + K_y)}{4}.   \label{FullHamiltonianSpecialPoint}
\end{aligned}
\end{equation}
The summation in \eqref{FullHamiltonianSpecialPoint} is a positive semidefinite matrix. As noted in Ref. \cite{richards2024exact}, for \(K_x, K_y \neq 0\), the ground state at these points is two-fold degenerate with energy \(E_0 = - N_c (K_x + K_y)/4\) and wavefunctions
\begin{equation}
\begin{aligned} 
|\Psi_G^* \rangle = \begin{Bmatrix} ( |- x \rangle \otimes |- y \rangle )^{\otimes N_c}, \\ ( |- y \rangle \otimes |- x \rangle )^{\otimes N_c} \end{Bmatrix}. 
\label{FullHamiltonianSpecialPointGSWavefunctions}
\end{aligned}
\end{equation}
It is clear that the exactly solvable points are contained within a phase characterized by a staggered chirality,
as shown by the pink stars in Figs. \ref{Alpha15PhaseDiagram} and \ref{Alpha5Alpha30PhaseDiagram}.
The ground state exhibits a soliton in OBCs which interpolates between the two degenerate ground states found in PBCs, as is typical of topological solitons \cite{IsotropicKitaevUnderFieldPaper, dauxois2006physics}.

\section{Large Unit-Cell Phases and Higher-Order Interactions\label{luphasediscussion}}
\begin{figure}[!ht]
    \centering
        \centering    
        \begin{centering}      \includegraphics[width=0.86\columnwidth]{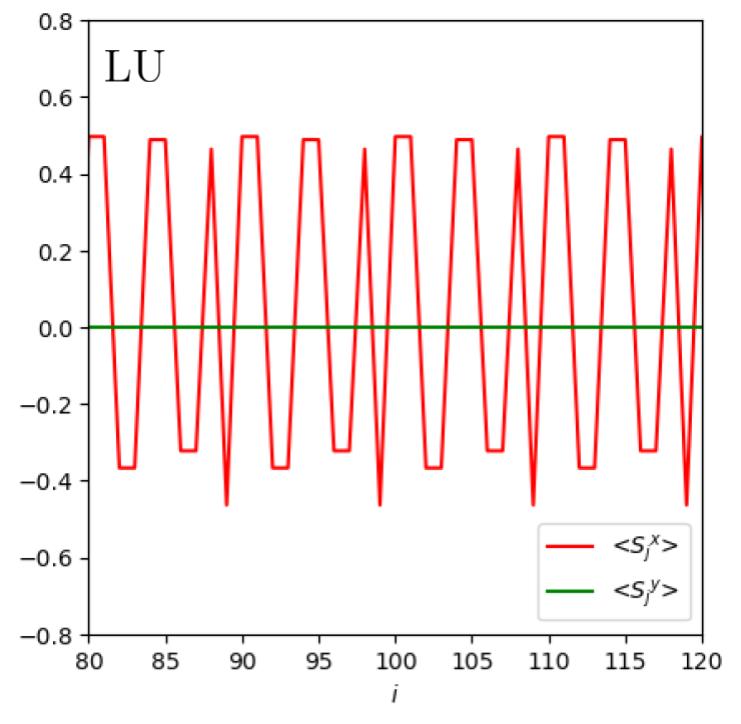}
        \end{centering}
  \caption{
  The on-site magnetizations \(\langle S_i^x \rangle\), \(\langle S_i^y \rangle\) versus position \(i\) in the middle of the chain in the LU phase, obtained from \(N = 200\) site DMRG with OBCs for a fixed \(\alpha = 15^{\circ}\), \(h = 0.57\), and \(\phi_{xy} = 0^{\circ}\).}
\label{Alpha15LUPhaseMagnetizations}
\end{figure}

As indicated by the blue triangle area of Fig. \ref{Alpha15PhaseDiagram}, DMRG at higher fields near \(h_y = 0\) seems to reveal a small window of other phases between the \(\mathcal{M}\) phase and the \(\mathrm{PS}\), which are denoted by \(\mathrm{LU}\). The on-site magnetizations at a point within a \(\mathrm{LU}\) phase is shown in Fig. \ref{Alpha15LUPhaseMagnetizations}, which reveals a periodicity of 10 sites. This result implies that the \(\mathrm{LU}\) phases carry larger unit cell orders.

To understand the appearance of the \(\mathrm{LU}\) phases in the higher \(h_x\) regime around \(h_y \approx 0\), we obtain higher order terms in the perturbation theory.
Upon setting \(h_y = 0\) in the model given by Eq. \eqref{FullHamiltonian}, and deriving a fourth-order effective Hamiltonian in the expansion $\frac{K_x}{2} > h_x \gg K_y $, we find the result
\begin{equation}
\begin{aligned} \boldsymbol{\mathcal{H}}_{\mathrm{eff}}^{\mathrm{4}^{\mathrm{th}}}[\mathbf{h} = h_x \mathbf{\hat{x}}]  \simeq  \sum_{j = 1}^{N_c} \Bigg\{  \tilde{K}_T  \tau^z_{j} \tau^z_{j + 1}  + \tilde{K}_{T,2} \tau^z_{j} \tau^z_{j + 2}     \Bigg\},
\label{FourthOrderEffectiveHamiltonianZerohy}
\end{aligned}
\end{equation}
where  
\begin{equation}
\begin{aligned}     &\tilde{K}_T \equiv  \bigg( \frac{h_x^2 }{8  K_x( K_x^2 - 4h_x^2)}  \bigg) K_y^2, \\   &\tilde{K}_{T,2} \equiv \bigg( \frac{ (9 K_x^4  - 28 K_x^2 h_x^2 + 32 h_x^4) h_x^2}{128 K_x^3 ( K_x^2 - 4h_x^2)^3} \bigg) K_y^4.
\label{FourthOrderEffectiveHamiltonianZerohyConstants}
\end{aligned}
\end{equation} \(\boldsymbol{\mathcal{H}}_{\mathrm{eff}}^{\mathrm{4}^{\mathrm{th}}}\) is the one-dimensional axial next-nearest neighbor Ising (ANNNI) chain with antiferromagnetic couplings. The ground state of the effective \(\tau\)-model \(\boldsymbol{\mathcal{H}}_{\mathrm{eff}}^{\mathrm{4}^{\mathrm{th}}}\) is antiferromagnetic when \(\tilde{K}_{T} > 2 \tilde{K}_{T,2}\) and is thus part of the \(\mathcal{M}\) phase. 

However, when \(\tilde{K}_{T} < 2 \tilde{K}_{T,2}\), the system transitions to a new phase known as an `antiphase'. Here, the ground state becomes four-fold degenerate \cite{ANNNIPaper,domb1960theory,liebmann1986statistical,selke1988annni, chakrabarti2008quantum} with states such as \(\bigotimes_{j = 1}^{N_c/4} \big( |a_{4j - 3} \rangle \otimes |a_{4j - 2} \rangle \otimes |b_{4j - 1} \rangle \otimes |b_{4j} \rangle  \big)\)  where the next-nearest neighbour spins are antialigned \cite{ANNNIPaper}. 
Written in terms of original spin operators, these four states are
\begin{equation}
\begin{aligned} |\Psi_G^{\mathcal{A}} \rangle = \begin{Bmatrix}
    \big( |\uparrow \downarrow  \uparrow \downarrow  \downarrow \uparrow \downarrow \uparrow \big)^{\otimes N_c/4}, \\  \big( |\uparrow \downarrow \downarrow \uparrow \downarrow \uparrow  \uparrow \downarrow \rangle  \big)^{\otimes N_c/4}, \\ \big( | \downarrow \uparrow  \uparrow \downarrow \uparrow \downarrow  \downarrow \uparrow   \rangle \big)^{\otimes  N_c/4}, \\ \big(| \downarrow \uparrow  \downarrow \uparrow   \uparrow \downarrow \uparrow \downarrow  \rangle \big)^{\otimes N_c/4},   
\end{Bmatrix},
\label{APhaseGroundState}
\end{aligned}
\end{equation}
which carry 8-site periodicity. When \(\tilde{K}_{T} = 2 \tilde{K}_{T,2}\), the ground state is infinitely degenerate with the degeneracy equal to \( ((1 + \sqrt{5})/2)^{N_c} \) \cite{ANNNIPaper, selke1988annni, barber1982hamiltonian, chakrabarti2008quantum}. By Eq. \eqref{FourthOrderEffectiveHamiltonianZerohyConstants}, this point occurs at the field value  
\begin{equation}
\begin{aligned} h_{x,c_2} = \frac{K_x}{4} \sqrt{\frac{ 16 K_x^2 - 7 K_y^2  -  K_y \sqrt{  128 K_x^2 - 23 K_y^2  }}{4 K_x^2 - K_y^2}},
\label{PerturbationTheoryMPhaseBoundary}
\end{aligned}
\end{equation} 
which is a critical field for the transition from the \(\mathcal{M}\) phase to a phase with 8-site unit cell order within the strong anisotropic regime. The dashed blue line in the inset of Fig. \ref{hxEqual0ClassicalPhaseBoundary} shows the critical field \(h_{x,c_2}\) for reasonable \(\alpha\) within the strong anisotropic regime (we choose \(0 \leq \alpha \leq 15^{\circ}\) - values reasonable in accordance to fourth order perturbation theory). For comparison, the classical phase boundary given by the black line is also shown. The point where we observe the on-site magnetizations with 10-site order in Fig. \ref{Alpha15LUPhaseMagnetizations} is shown by the gold star. 

The 8-site order found from \(\boldsymbol{\mathcal{H}}_{\mathrm{eff}}^{\mathrm{4}^{\mathrm{th}}}\) for \(h_x > h_{x,c_2}\) is different from the 10-site order found from DMRG. This is due to the neglected higher order terms. Despite the clear discrepancy, the perturbation theory reveals the transition from the \(\mathcal{M}\) phase to a \(\mathrm{LU}\) phase. It also reveals that the \(\mathcal{M}\) phase is more extended in the fourth order effective \(\tau\)-model in comparison to the classical analysis, as seen by comparing the dashed blue and black lines in Fig. \ref{Alpha15LUPhaseMagnetizations}. However, both analyses do not align with the actual critical field of the original \(\sigma\)-model where the \(\mathcal{M}\) phase closes.   

It is possible that a series of \(\mathrm{LU}\) phases including incommensurate order may emerge as the field strength \(h\) and angle \(\phi_{xy}\) vary. This is consistent with DMRG studies where the convergence of the ground states is difficult to observe within \(\mathrm{LU}\) and the white circle regime shown in Fig. \ref{Alpha15PhaseDiagram}. Determining the precise nature of the \(\mathrm{LU}\) phases and obtaining their phase boundaries using DMRG techniques are left for future work.

\section{Summary and discussion\label{summary}}
In summary, we have studied anisotropic spin-1/2 Kitaev chains with bond strengths \(K_x > K_y\) under an applied external magnetic field \(\mathbf{h}\) through a combination of numerical and analytical techniques. Given that the model is exactly solvable under a transverse magnetic field \(\mathbf{h} = (0, 0, h_z)\) and is found to directly enter the PS at any finite \(h_z\), we focus on longitudinal fields \(\mathbf{h} =  (h_x, h_y, 0)\). Using the finite DMRG method, we obtained the phase diagram in the \((h_x, h_y)\) plane and have found the occurrence of several nontrivial phases - \(\mathcal{M}\) which occurs at both low and higher fields where \(h_x > h_y \geq 0\), and \(\mathcal{C}\) and \(\mathcal{B}\) which occur at higher \(h_x\) fields with a smaller but sizable \(h_y\) component. We have also found the emergence of \(\mathrm{LU}\) in a small phase space before the system polarizes from the \(\mathcal{M}\) phase at higher \(h_x\) fields around \(h_y = 0\). The \(\mathcal{M}\) and \(\mathcal{C}\) phases were found to be field-induced ordered phases that possess magnetic unit cells of 4-sites and 6-sites, respectively. The \(\mathcal{B}\) phase was found to be a soliton phase that was also previously observed in the isotropic chain where \(K_x = K_y\). The \(\mathrm{LU}\) phases appear to be phases that possess larger magnetic unit cells, with DMRG indicating an ordering with 10-site unit cells.  

To understand the nature of phases \(\mathcal{M}\) and \(\mathcal{C}\), we derived an effective low-energy model including third and fourth order terms in the strong anisotropic regime. The effective model revealed that the \(\mathcal{M}\) phase is characterized by a staggered magnetization and it occurs due to transverse Ising interactions between the unit cells enabled by the external field. Study of the classical phase diagram revealed that as the ratio of the bond strengths \(K_y/K_x\) increases, the \(\mathcal{M}\) phase shrinks until it vanishes in the isotropic limit, in agreement with DMRG studies. The effective model also revealed that the \(\mathcal{C}\) phase is characterized by a uniform chirality in addition to 6-site staggered order, and it occurs due to Dzyaloshinskii-Moriya interactions between the unit cells enabled by higher external fields. The \(\mathcal{B}\) phase was explained through exactly solvable points of the model at the higher field values \(\mathbf{h} =  (K_x/2, K_y/2, 0)\), where the two-fold degenerate ground state was exactly obtained. 
The \(\mathrm{LU}\) phases in the higher \(h_x\) regime around \(h_y = 0\) were understood with a fourth order effective model, which revealed interactions between next-nearest neighbour unit cells that dominate in a small regime of higher \(h_x\) fields beyond the point where the \(\mathcal{M}\) phase closes. 
A series of \(\mathrm{LU}\) phases including incommensurate order may exist as well. The precise nature of the \(\mathrm{LU}\) phases is not well understood, however, and further analytical and numerical investigations of them is left for future work.

Other future work involves finding candidate materials which shows the physics of the anisotropic Kitaev chain. Recently,  CoNb\(_2\)O\(_6\) has been shown to represent a rare one-dimensional Kitaev chain with significant bond-dependent ferromagnetic Kitaev and antiferromagnetic \(\Gamma\) interactions. The inclusion of these interactions gives rise to Ising anisotropy and domain wall motion at zero applied transverse magnetic field \cite{churchill2024transforming}, which is consistent with what is observed in experiments such as inelastic neutron scattering and THz spectroscopy \cite{churchill2024transforming, morris2021duality, coldea2010quantum}. Finding an antiferromagnetic sister material which shows anisotropic Kitaev physics would be of considerable interest. 

It is noteworthy that the field-induced intermediate phases between antiferromagnetic (AFM) order and polarized states in $\alpha$-RuCl$_3$ \cite{kasahara2018thermal}, a prime candidate among Kitaev materials, have garnered significant attention. While debates persist regarding the existence and nature of these intermediate phases \cite{Rouso2024RoPP}, theoretical studies have identified large unit-cell phases in the classical Kitaev-$\Gamma$ models \cite{Liu2021PRR, Rayyan2021}, which appear as intermediate phases between the AFM and polarized states when the magnetic field is applied along the c-axis, perpendicular to the honeycomb plane \cite{Chern2020PRR}. The AFM $\Gamma$ interaction renders the c-axis a hard axis \cite{Plumb2014PRB}, meaning the system is resistant to polarization along this direction. This behavior is reminiscent of the hard axis found in the strong-bond field of the anisotropic Kitaev chain. Understanding the mechanism behind such large unit-cell orders in the 2D Kitaev-$\Gamma$ model, similar to the \(\mathrm{LU}\) phases found in the anisotropic Kitaev chain, could be an interesting subject for future research. \\

\section*{Acknowledgments}
This work is supported by the NSERC Discovery Grant No. 2022-04601. H.Y.K acknowledges support from the Canada Research Chairs Program and 
the Simons Emmy Noether fellowship of the Perimeter Institute,
supported by a grant from the Simons Foundation (1034867, Dittrich).
Computations were performed on the Niagara supercomputer at
the SciNet HPC Consortium. SciNet is funded by: the
Canada Foundation for Innovation under the auspices of
Compute Canada; the Government of Ontario; Ontario
Research Fund - Research Excellence; and the University
of Toronto.


%

\appendix

\section{DMRG results for other anisotropic strengths \label{MoreNumericalDataSection}}

\subsubsection{Phase diagram in $h_x-h_y$ phase space for different \(\alpha\) values\label{MoreAlphaPhaseDiagramsSection} } 

\begin{figure}[!ht]
    \centering
        \centering
        \begin{centering}        \includegraphics[width=0.95\columnwidth]{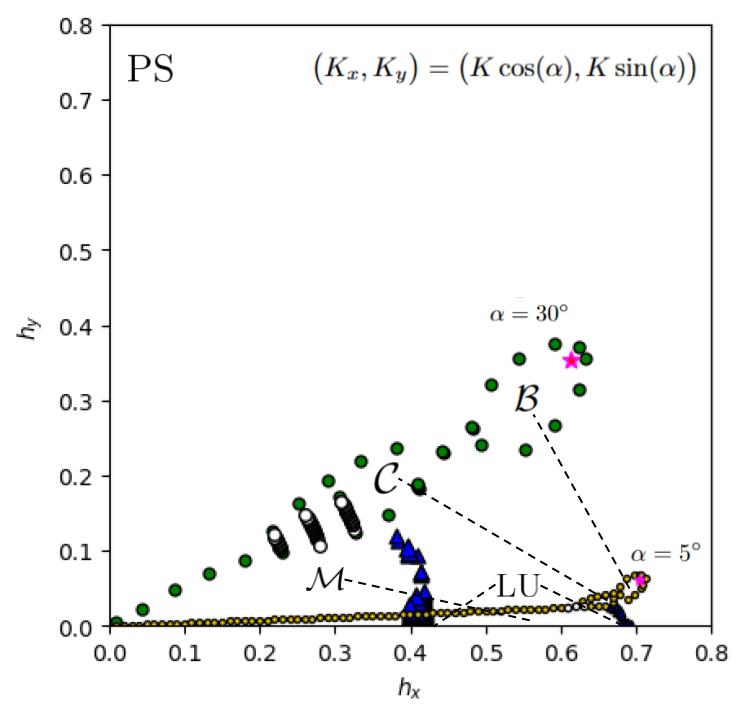}    
        \end{centering}
  \caption{The phase diagram for the anisotropic chains for \(\alpha = 5^{\circ}\) (solid gold points) and \(\alpha = 30^{\circ}\) (solid green points) in the \((h_x, h_y)\) plane, obtained from DMRG for \(N = 200\) sites, with PBCs. Blue triangles denote transition points from the \(\mathcal{M}\) phase to other smaller phases denoted by \(\mathrm{LU}\) occurring before the system reaches the \(\mathrm{PS}\).  White circles denote a small region which separates the $\mathcal{M}$ and $\mathcal{C}$ phases (see the main text for details). Pink stars mark the exactly solvable points for the \(\mathcal{B}\) phase discussed in Sec. \ref{exactlysolvablepointsection}.}
\label{Alpha5Alpha30PhaseDiagram}
\end{figure}

\renewcommand{\theequation}{A.\arabic{equation}}
\setcounter{equation}{0}
We show the phase diagrams for \(\alpha = 30^{\circ}\) and \(5^{\circ}\) in Fig. \ref{Alpha5Alpha30PhaseDiagram}.  
The latter is closer to the perturbative regime described by the effective \(\tau\)-models. The field induced phases \(\mathcal{M}\), \(\mathcal{C}\), and \(\mathcal{B}\) are found for all cases presented. The \(\mathcal{C}\) and \(\mathcal{B}\) phases are present only when $h_y$ is finite. The \(\mathcal{M}\) phase shrinks as $\alpha$ increases, as we discussed in the main text.

\subsubsection{Unit cell correlators \(\langle \xi_{\sigma,j} \rangle\)  \label{UnitCellCorrelatorsSection} } 
We show below in Fig. \ref{CPhaseUnitCellUniformChiralities} the unit cell correlators \(\langle \xi_{\sigma,j} \rangle\) defined in Eq. \eqref{CPhaseUniformChirality} as a function of the unit cell \(j\) for OBCs. We only show values for unit cells in the middle of the chain. As shown, the \(\mathcal{C}\) phase posses both uniform and staggered chirality, but the staggered chirality is a trivial effect due to the field.  

\begin{figure}[!ht]
    \centering
        \centering
        \begin{centering}        \includegraphics[width=0.88\columnwidth]{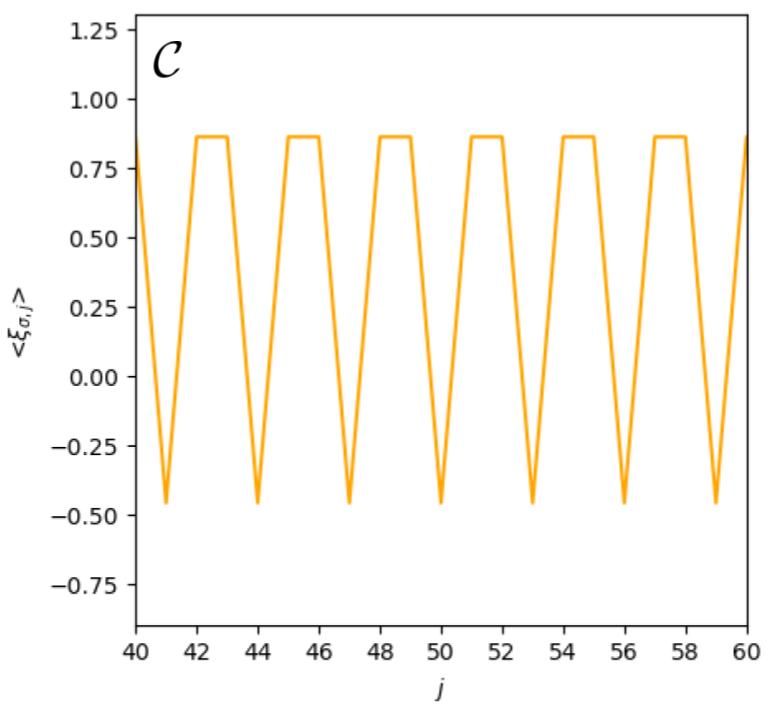}    
        \end{centering}
  \caption{The unit cell correlators \(\langle \xi_{\sigma,j} \rangle\) versus unit cell \(j\) in the middle of the chain in the \(\mathcal{C}\) phase, obtained from \(N = 200\) site DMRG with OBCs for a fixed \(\alpha = 15^{\circ}\), \(h = 0.55\), and \(\phi_{xy} = 12^{\circ}\).}
\label{CPhaseUnitCellUniformChiralities}
\end{figure}

\section{Linear Spin Wave Theory \label{linearspinwavetheoryappsection}} 

\renewcommand{\theequation}{B.\arabic{equation}}
\setcounter{equation}{0}

\subsubsection{\(\mathcal{M}\) phase along \(h_{y} = 0\) \label{OrderedPhaseSpinWaveTheorySection} } 
Given the classical spin configuration Eq. \eqref{ClassicalMPhaseSpinConfiguration}, we define a new magnetic unit cell indexed by \(l\) which contains sites \(2j-1,A/B \) and \( 2j,A/B\) as such:
\begin{equation}
\begin{aligned} \mathbf{S}_{l,\tilde{A}} = \pm S \mathbf{\hat{x}}, \,  \mathbf{S}_{l,\tilde{B}} = \mp S \mathbf{\hat{x}}, \,  \mathbf{S}_{l,\tilde{C}} = \mp S \mathbf{\hat{x}}, \,  \mathbf{S}_{l,\tilde{D}} = \pm S \mathbf{\hat{x}}, \\ l = 1, ..., N_c/2. 
\label{ClassicalMPhaseSpinConfigurationRewritten}
\end{aligned}
\end{equation}
The model in this case can be rewritten as  
\begin{equation}
\begin{aligned} \mathcal{H}  = \sum_{l = 1}^{N_l} \bigg[  K_x S_{l,\hat{A}}^x S_{l,\hat{B}}^x + K_y S_{l,\hat{B}}^y S_{l,\hat{C}}^y +  K_x S_{l,\hat{C}}^x S_{l,\hat{D}}^x \\ + K_y S_{l,\hat{D}}^y S_{l + 1,\hat{A}}^y +  h_x (S^x_{l,\hat{A}} + S^x_{l,\hat{B}} + S^x_{l,\hat{C}} + S^x_{l,\hat{D}})\bigg], \\ N_l \equiv \frac{N_c}{2}. 
\label{FullHamiltonianRewritten}
\end{aligned}
\end{equation}
We choose site dependent frames in the Holstein-Primakoff transformation so that \(\mathbf{\hat{z}}_{k}\) is aligned with the local ordering direction,
\begin{equation}
\begin{aligned} (\mathbf{\mathbf{\hat{x}}}_{l,\tilde{A}}, \, \mathbf{\mathbf{\hat{y}}}_{l,\tilde{A}}, \, \mathbf{\mathbf{\hat{z}}}_{l,\tilde{A}}) = (\mathbf{\mathbf{\hat{y}}}, \, \pm \mathbf{\mathbf{\hat{z}}}, \, \pm \mathbf{\mathbf{\hat{x}}}), \\ (\mathbf{\mathbf{\hat{x}}}_{l,\tilde{B}}, \, \mathbf{\mathbf{\hat{y}}}_{l,\tilde{B}}, \, \mathbf{\mathbf{\hat{z}}}_{l,\tilde{B}}) = (\mathbf{\mathbf{\hat{y}}}, \, \mp \mathbf{\mathbf{\hat{z}}}, \, \mp \mathbf{\mathbf{\hat{x}}}), \\ (\mathbf{\mathbf{\hat{x}}}_{l,\tilde{C}}, \, \mathbf{\mathbf{\hat{y}}}_{l,\tilde{C}}, \, \mathbf{\mathbf{\hat{z}}}_{l,\tilde{C}}) = (\mathbf{\mathbf{\hat{y}}}, \, \mp \mathbf{\mathbf{\hat{z}}}, \, \mp \mathbf{\mathbf{\hat{x}}}), \\ (\mathbf{\mathbf{\hat{x}}}_{l,\tilde{D}}, \, \mathbf{\mathbf{\hat{y}}}_{l,\tilde{D}}, \, \mathbf{\mathbf{\hat{z}}}_{l,\tilde{D}}) = (\mathbf{\mathbf{\hat{y}}}, \, \pm \mathbf{\mathbf{\hat{z}}}, \, \pm \mathbf{\mathbf{\hat{x}}}), \\ j = 1, ..., N_c/2.  
\label{LocalOrderingMPhase}
\end{aligned}
\end{equation}

We define four bosons pertaining to the four sublattice sites \(A\), \(B\), \(C\) and \(D\), whose creation and annihilation operators are \(a^{\dagger}/a\), \(b^{\dagger}/b\), \(c^{\dagger}/c\) and \(d^{\dagger}/d\), respectively. In this case, we can write the Holstein-Primakoff transformation \cite{holstein1940field,rau2018pseudo} as
\begin{equation}
\begin{aligned} \mathbf{S}_{l,\hat{A}/\hat{B}/\hat{C}/\hat{D}} =  \sqrt{\frac{S}{2}} ( (a/b/c/d)_{l}^{\dagger} + (a/b/c/d)_{l}) \mathbf{\hat{x}}_{l,\hat{A}/\hat{B}/\hat{C}/\hat{D}} \\ + i \sqrt{\frac{S}{2}} ( (a/b/c/d)_{l}^{\dagger} - (a/b/c/d)_{l})  \mathbf{\hat{y}}_{l,\hat{A}/\hat{B}/\hat{C}/\hat{D}} \\ + (S - (a/b/c/d)_{l}^{\dagger} (a/b/c/d)_{l} ) \mathbf{\hat{z}}_{l,\hat{A}/\hat{B}/\hat{C}/\hat{D}}, \\ l = 1, ..., N_c/2.  
\label{HolsteinPrimakoffTransformation}
\end{aligned}
\end{equation}
We sub Eq. \eqref{LocalOrderingMPhase} into the Holstein-Primakoff transformation,
and then sub the resulting transformation into the rewritten Hamiltonian Eq. \eqref{FullHamiltonianRewritten}. In discarding linear terms and and looking no further than quadratic terms in the boson operators, we find two contributions:

\begin{equation}
\begin{aligned} \mathcal{H} = E_{\mathrm{cl}} + \mathcal{H}^{(2)}, 
\label{HamiltonianClassicalQuadratic}
\end{aligned}
\end{equation}
where 
\begin{equation}
     E_{\mathrm{cl}} = - N_c S (S + 1) K_x   \label{MPhaseClassicalEnergy}
\end{equation}
and \(\mathcal{H}^{(2)}\) contains the quadratic terms in the boson operators, which can be expressed in terms of the momentum space operators 
\begin{equation}
     \mathbf{\hat{a}}_{q} = \frac{1}{\sqrt{N_l}} \sum_{l} e^{-i ql } \mathbf{\hat{a}}_{l},  \label{InverseFourierTransform}
\end{equation}
where \(\mathbf{\hat{a}}_{q/l} \equiv \begin{Bmatrix}a_{q/l} & b_{q/l} & c_{q/l} & d_{q/l} \end{Bmatrix}^T\). We define the Nambu spinors
\begin{equation}
     \hat{A}_{q} = \begin{Bmatrix}
         \mathbf{\hat{a}}_{q} & \mathbf{\hat{a}}_{-q}^{\dagger} 
     \end{Bmatrix}^{T}, \, \, [ \hat{A}_{q}, \hat{A}_{p}^{\dagger} ] = \delta_{q,p} \delta_{l,r} \begin{pmatrix}
         + \mathbbm{1} & 0 \\ 0 & -\mathbbm{1}
     \end{pmatrix} = \delta_{q,p} 
     \sigma^z,  \label{NambuSpinors}
\end{equation}
where \(l\) and \(r\) are real space positions, and with these, the result of \(\mathcal{H}^{(2)}\) in momentum space is found to be 
\begin{equation}
\begin{aligned} \mathcal{H}^{(2)} = \frac{S}{2} \sum_{\mathbf{q}} \hat{A}_{q}^{\dagger} \boldsymbol{\mathrm{H}} \hat{A}_{q},
\label{QuadraticHamiltonianDiagonalized}
\end{aligned}
\end{equation}
where \(\boldsymbol{\mathrm{H}}\) is a \(8 \times 8\) Hermitian matrix with nonzero elements
\begin{equation}
\begin{aligned} \mathbf{H}_{11}  = \mathbf{H}_{44} = \mathbf{H}_{55}  = \mathbf{H}_{88} = K_x \pm \frac{h_x}{S} , \\ \mathbf{H}_{22}  = \mathbf{H}_{33} = \mathbf{H}_{66}  = \mathbf{H}_{77} =  K_x \mp \frac{h_x}{S} , \\   \mathbf{H}_{23} =  \mathbf{H}_{27}  = \mathbf{H}_{36}  = \mathbf{H}_{67}  = \frac{K_y}{2}, \\    \mathbf{H}_{14}  =  \mathbf{H}_{18} = \mathbf{H}_{54} = \mathbf{H}_{58} = \frac{K_y}{2} e^{ik}.
\label{MPhaseQuadraticElements}
\end{aligned}
\end{equation}
After a bosonic Bogoliubov transformation, \(\mathcal{H}^{(2)} \) \eqref{QuadraticHamiltonianDiagonalized} can be brought into the canonical form 
\begin{equation}
\begin{aligned} \mathcal{H}^{(2)} = S \sum_{q,\alpha } \omega_{q,\alpha} \bigg( \gamma_{q,\alpha}^{\dagger} \gamma_{q,\alpha} + \frac{\mathbbm{1}}{2} \bigg),
\label{QuadraticHamiltonianFourierTransformed}
\end{aligned}
\end{equation}
where \(\alpha\) in this case indexes over 4 different magnon branches. From Eq. \eqref{QuadraticHamiltonianFourierTransformed}, we identify the quantum zero-point energy  
\begin{equation}
\begin{aligned} E_{\mathrm{qu}} = \frac{S}{2} \sum_{q,\alpha } \omega_{q,\alpha}.  
\label{MPhaseZeroPointEnergy}
\end{aligned}
\end{equation}
The ground state energy within linear spin wave theory is the sum of the classical energy \(E_{\mathrm{cl}}\) and the zero point energy \(E_{\mathrm{qu}}\), i.e.
\begin{equation}
\begin{aligned} E_{0} = E_{\mathrm{cl}} +  E_{\mathrm{qu}}.  
\label{TotalZeropointEnergy}
\end{aligned}
\end{equation}
\\
The magnon branches can be found by diagonalizing \(\sigma^z \mathbf{H}\), which gives us the eigenvalues in \(\pm\) pairs \(\{\omega_{q,\alpha}, - \omega_{-q,\alpha}\}\) \cite{LinearSpinWaveTheoryPaper}. The 4 distinct branches in this case are given by Eq. \eqref{MPhaseMagnonBranches} in the main text.
\\
\subsubsection{Polarized state \label{PolarizedStateSpinWaveTheorySection}} 
Given the classical spin configuration Eq. \eqref{ClassicalAntialignedState},
we choose site dependent frames in the Holstein-Primakoff transformation so that \(\mathbf{\hat{z}}_{j}\) is aligned with the spins, i.e.
\begin{equation}
\begin{aligned} (\mathbf{\mathbf{\hat{x}}}_{j,A/B}, \, \mathbf{\mathbf{\hat{y}}}_{j,A/B}, \, \mathbf{\mathbf{\hat{z}}}_{j,A/B}) = (-  \mathbf{\hat{h}}_{\perp}, \,  \mathbf{\mathbf{\hat{z}}}, \, - \mathbf{\hat{h}}), \\ j = 1, ..., N_c,  
\label{LocalOrderingAntialignedState}
\end{aligned}
\end{equation}
where \(  \mathbf{\hat{h}}_{\perp} = \partial_{\phi_{xy}} \mathbf{\hat{h}} \).

We sub Eq. \eqref{LocalOrderingAntialignedState} into 
the Holstein-Primakoff transformation and then sub the resulting transformation into the Hamiltonian. 
In discarding linear terms and looking no further than quadratic terms in the boson operators, we again find two contributions given by Eq. \eqref{HamiltonianClassicalQuadratic}, where
\begin{equation}
     E_{\mathrm{cl}} = N_c S \bigg( S + \frac{1}{2}\bigg) \bigg[   K_x   \cos^2(\phi_{xy})  +  K_y   \sin^2(\phi_{xy})  -   \frac{2 h}{S}  \bigg]   \label{AntialignedStateClassicalEnergy}
\end{equation}
and the quadratic component \(\mathcal{H}^{(2)}\) can be written in the form of Eq. \eqref{QuadraticHamiltonianDiagonalized}, but this time \(\mathbf{\hat{a}}_{q/j} \equiv \begin{Bmatrix}a_{q/j} & b_{q/j} \end{Bmatrix}^T\) in the Nambu spinors given by Eq. \eqref{NambuSpinors}, and \(\mathbf{H}\) is now a \(4 \times 4\) Hermitian matrix with nonzero elements 
\begin{equation}
\begin{aligned} \mathbf{H}_{11} = \mathbf{H}_{22}  = \mathbf{H}_{33} = \mathbf{H}_{44 } = \\  - K_x \cos^2 (\phi_{xy})  -  K_y  \sin^2 (\phi_{xy})  + \frac{h}{S} , \\ \mathbf{H}_{12} = \mathbf{H}_{14}  = \mathbf{H}_{32} = \mathbf{H}_{34} = \\  \frac{1}{2}  K_x \sin^2(\phi_{xy}) + \frac{1}{2} K_y \cos^2(\phi_{xy})   e^{ik}.
\label{AntialignedStateQuadraticElements}
\end{aligned}
\end{equation}
We perform the bosonic Bogoliubov transformation to bring \( \mathcal{H}^{(2)}\) into the canonical form given by Eq.  \eqref{QuadraticHamiltonianFourierTransformed}, where \(\alpha\) now indexes over 2 different magnon branches, which are again found by diagonalizing \(\sigma^z \mathbf{H}\). The resulting magnon branches are found to be \cite{gordon2023exploring}
\begin{equation}    
 \begin{aligned}
 \omega_{\pm}(k)  =  \sqrt{ \gamma \big( \gamma  \pm   |\xi(k)| \big) },
 \end{aligned}
 \label{AntialignedStateMagnonBranches}
 \end{equation}
 where 
 \begin{equation}
 \begin{aligned}
 \gamma & \equiv  K_x \cos^2(\phi_{xy}) +  K_y \sin^2(\phi_{xy}) - \frac{h}{S}, \\ 
 \xi(k) & \equiv K_x \sin^2(\phi_{xy}) + K_y \cos^2(\phi_{xy}) e^{ik}.
\label{AntialignedStateMagnonBranchesConstants}
\end{aligned}
\end{equation}
When \(h_y = 0\) (i.e. \(\phi_{xy} = 0^{\circ}\)), the branches are identical to \(\omega_{(-),\pm}\) given by Eq. \eqref{MPhaseMagnonBranches} in the main text.

\end{document}